 \documentclass[prc,amsmath,amsfonts,showpacs,eqsecnum,twocolumn]{revtex4}

 \usepackage{graphicx}  
 \usepackage{pstcol}    
 
\begin{document}
\hyphenation{Rijken}
\hyphenation{Nijmegen}
 
\title{
    ESC NN-Potentials in Momentum Space \\ 
    I. PS-PS Exchange Potentials }
\author{Th.A.\ Rijken}
\author{H.\ Polinder}
\affiliation{Institute for Theoretical Physics Nijmegen, University of Nijmegen,
         Nijmegen, The Netherlands } 
\author{ J.\ Nagata}                             
\altaffiliation{Present address: Kyushu International University, Fukuoka 805-8512, Japan}                     
\affiliation{Venture Business Laboratory, Hiroshima University, Kagamiyama 2-313, 
         Higashi-Hiroshima, Japan } 
 
\date{version of: \today}

\begin{abstract}                                       
A momentum space representation is derived for the Nijmegen 
Extended-Soft-Core (ESC) interactions. 
The partial wave projection of this representation is carried through,
in principle for Two-Meson-Exchange (TME) in general. 
Explicit results for the momentum space partial wave $NN$-potentials 
from PS-PS-Exchange are given.              
 
\end{abstract}

\maketitle

\section{Introduction}                                     
\label{sec:1}
 
For nucleon-nucleon \cite{Rij93,RS96a,RS96b,SR97}  and 
hyperon-nucleon \cite{Rij99} scattering
we have shown that the Extended-Soft-Core (ESC) models for baryon-baryon 
interactions give an excellent description of the NN and YN data. 
So far, the applications of the ESC-model has been in configuration
space. In two papers we give a momentum space representation of the
ESC-potentials. Also, we describe a fit to the pp- and np-data for 
$0 \le T_{lab} \leq 350$ MeV, having $\chi^2/N_{data}=1.15$. Here, we 
used 20 physical parameters, being coupling constants and cut-off masses.

In synopsis, an account of a modern theoretical basis for the soft-core
interactions has been given in \cite{Rij93}. Starting from the so called
Standard Model and integrating out the heavy quarks one arrives at an effective 
QCD for the u,d,s quarks. The generally accepted scenario is now that the
QCD-vacuum becomes unstable for momenta transfers for which 
$Q^2 \leq \Lambda_{\chi SB}^2 \approx 1$ GeV$^2$ \cite{Geo93}, causing 
spontaneous chiral-symmetry breaking ($\chi$SB).
The vacuum goes through a phase-transition, generating constituent quark masses
via $\langle 0|\bar{\psi}\psi|0\rangle \neq 0$, and reducing the gluon coupling
$\alpha_s$. Viewing the pseudo-scalar mesons $\pi$ etc. as the Nambu-Goldstone bosons 
originating from the spontaneous $\chi SB$, makes it natural to assume quark
dressing by pseudo-scalar mesons, and also other types of mesons in general.
In this context, baryon-baryon interactions are described naturally by meson-exchange
using form factors at the meson-baryon vertices. 
As a working hypothesis, we restrict ourselves for low energy scattering 
to a treatment of the the hadronic-phase. 
Integrating out the heavy mesons and baryons using
a renormalization procedure a la Wilson \cite{Pol84}, we restrict ourselves to
mesons with $M \leq 1$ GeV$/c^2$, arriving at an so-called 
{\it effective field theory} as the proper arena to describe low energy hadron 
scattering.  This is the general physical basis for the Nijmegen soft-core models.

In this paper, hereafter referred to as I, a representation for the 
ESC-interactions in momentum space is presented, which is very elegant and 
useful for applications in  momentum space computations. 

{\it
We solve basically the problem of finding a suitable representation 
of the TME-potentials in momentum space in general, and in particularly with
gaussian form factors.
As is apparant from the configuration 
space representation, given in \cite{Rij91,RS96a,RS96b}, we have to evaluate forms like
\[
 \frac{2}{\pi}\int_0^\infty d\lambda\ \tilde{F}(\mbox{\boldmath $\Delta$}^2,\lambda)\ 
 \tilde{G}((\Delta-{\bf k})^2,\lambda)\ ,
\]
which at first sight means an extra numerical integral, besides the usual convolutive
integration over the $\mbox{\boldmath $\Delta$}$-momentum.
Here, $\tilde{F}$ and $\tilde{G}$ contain the couplings, gaussian form factors,
and momentum space Yukawa functions for the two exchanged mesons.
Fortunately, the integrations over the $\lambda$-parameter can be carried 
through analytically, as we shall show in the sequel.
The representations obtained in this paper contain only two parameters,
henceforth called t and u, to be integrated over numerically. 
Moreover, the $(t,u)$-integration region is effectively over a finite domain, 
this in view of the exponentials $\exp(-m_1^2 t)$ and
$\exp(-m_2^2 u)$ in the integrands.
Also, the strong gaussian-like fall off in the off-shell region, typical for the soft-core 
interactions, is easily realized. To realize the latter property of the momentum space 
potentials is problematic, when these potentials have to be generated numerically 
from configuration space via Fourier transformation.
}

The partial wave projection formulas on the $LSJ$-partial wave basis 
are developed for the TME-potentials in momentum space. 
Like for the OBE-potentials in \cite{RKS91}, we expand the momentum space 
potentials in the Pauli-spinor invariants.                    
The partial wave basis is choosen according to the convention
of \cite{SYM57}.
Explicit formulas are worked out for general pseudoscalar-pseudoscalar (ps-ps) 
exchange.

In paper II \cite{RN00b}, the momentum space representation for the ESC 
Meson-Pair-Exchange (MPE) potentials is described. Here, we moreover present
a fit to the $NN$-data for $0 \leq T_{lab} \leq 350$ MeV. In this ESC-model
included are the contributions from OBE-, PS-PS-exchange, and MPE.

The contents of this paper are as follows. In section II\ we review
the definition of the ESC-potentials in the context of the
relativistic two-body equations, the Thompson, and Lippmann-Schwinger equation.
Here, we exploit the Macke-Klein \cite{Klein53} framework in Field-Theory.
For the Lippmann-Schwinger equation we introduce the usual potential forms
in Pauli spinor space. We include here the central ($C$), the spin-spin
($\sigma$), the tensor ($T$), the spin-orbit ($SO$), the quadratic
spin-orbit ($Q_{12}$), and the antisymmetric spin-orbit ($ASO$)
potentials. For TME-exchange, in the approximations made in \cite{RS96a,RS96b}
only the central, spin-spin, tensor, and spin-orbit potentials occur.
In section III\ the special representation for the ESC-interactions
in momentum space is developed and illustrated using some basic examples.
In section IV\ the general ps-ps exchange is presented.
In section V\ the projection on the plane wave spinor-invariants is carried through 
for general ps-ps exchange. The adiabatic terms, non-adiabatic, 
pseudovectorvertex, and off-shell $1/M$-corections are given explicitly.
In section VI\ the partial wave analysis is performed.                               
Finally in section VII\ the partial wave contributions from the adiabatic terms and
the mentioned $1/M$-corrections are given.

Appendix A contains some basic integrals which are employed in the paper.
In Appendix B a dictionary is given containing the results of the convolutive
integrations for the operators that occur in ps-ps exchange.
Appendix C contains the $LSJ$ partial wave matrix elements of several important operators. 
In Appendix D the Fourier connection between coordinate and momentum space 
is illustrated in particularly for the ps-ps
exchange tensor potential.

\section{Two-Body Integral Equations in Momentum Space}

\subsection{Relativistic Two-Body Equations}
We consider the nucleon-nucleon reactions 
\begin{eqnarray}
&&  N(p_{a},s_{a})+N(p_{b},s_{b}) \rightarrow
    N(p_{a'},s_{a'})+N(p_{b'},s_{b'}) 
\label{eq:20.1} \end{eqnarray}
with the total and relative four-momenta for the initial
and the final states
\begin{equation}
\begin{array}{lcl}
 P = p_{a} + p_{b} &,& P' = p_{a'} + p_{b'}\ , \\
 p = \frac{1}{2}(p_{a}-p_{b}) &,& p' = \frac{1}{2}(p_{a'}-p_{b'})\ ,
\end{array} 
\label{eq:20.2} \end{equation}
which become in the center-of-mass system (cm-system) for a and b
on-mass-shell
\begin{equation}
 P = ( W , {\bf 0}) \hspace{0.2cm} , \hspace{0.2cm} p = ( 0 , {\bf p})
 \hspace{0.2cm} , \hspace{0.2cm} p' = ( 0 , {\bf p}') \ .
\label{eq:20.3a} \end{equation}
In general, the particles are off-mass-shell in the Green-functions.
In the following of this section, the on-mass-shell momenta for the initial
and final states are denoted respectively by $p_{i}$ and $p_{f}$.
So, $p_{a}^{0}=E_{a}({\bf p}_i)=\sqrt{{\bf p}_i^{2}+M_{a}^{2}}$ and
$p_{a'}^{0}=E_{a'}({\bf p}_f)=\sqrt{{\bf p}_f^{2}+M_{a'}^{2}}$, and
similarly for b and b'.
Because of translation-invariance $P=P'$ and
$W=W'=E_{a}({\bf p}_i)+E_{b}({\bf p}_i)=E_{a'}({\bf p}_f)+E_{b'}({\bf p}_f)$.
The two-particle states we normalize in the following way
\begin{eqnarray}
  \langle {\bf p}_{1}',{\bf p}_{2}'|{\bf p}_{1},{\bf p}_{2}\rangle
  &=& (2\pi)^{3}2E({\bf p}_{1})
  \delta^{3}({\bf p}_{1}'-{\bf p}_{1})\cdot \nonumber\\
  && \times (2\pi)^{3}2E({\bf p}_{2})
  \delta^{3}({\bf p}_{2}'-{\bf p}_{2})\ .
\label{eq:20.3b} \end{eqnarray}
 
The relativistic two-body scattering-equation reads \cite{Feyn49,Schw51,SB51}
\begin{eqnarray}
    \psi(p,P) &=& \psi^{0}(p,P) + G(p; P) \nonumber\\
 && \times \int\!d^{4}p'\; I(p,p')\  \psi(p',P)\ ,
\label{BSeq} \end{eqnarray}
where $\psi(p,P)$ is a $4\times 4$-matrix in Dirac-space.
The contributions to the kernel $I(p,p')$ come from the two-nucleon
irreducible Feynman diagrams.
In writing (\ref{BSeq}) we have taken out an overall $\delta^4(P'-P)$-function
and the total four-momentum conservation is implicitly understood henceforth.
 
The two-particle Green function $G(p;P)$ in (\ref{BSeq}) is simply the
product of the free propagators for the baryons of line (a) and (b).
The nucleon and more general for any baryon the Feynman propagators are given 
by the well known formula
\begin{eqnarray}
    G^{(s)}_{\{\mu\},\{\nu\}}(p) &=& \int d^{4}x\ \langle 0|T(
    \psi^{(s)}_{\{\mu\}}(x)\bar{\psi}^{(s)}_{\{\nu\}}(0))|0\rangle\
    e^{i p\cdot x} \nonumber \\[0.2cm]
  & = & \frac{\Pi^{s}(p)}{p^{2}-M^{2}+i\delta}
\label{Greens} \end{eqnarray}
where $\psi^{(s)}_{\{\mu\}}$ is the free Rarita-Schwinger field
which describes the nucleon $(s=\frac{1}{2})$, the $\Delta_{33}$-resonance 
$(s=\frac{3}{2})$, etc. (see for example \cite{Car71}). 
For the nucleon, the only case considered in this paper,  $\{\mu\}= \emptyset$
and for e.g. the  $\Delta$-resonance $\{\mu\}= \mu$. For the rest of this 
paper we deal only with nucleons.\\
In terms of these one-particle Green-functions
the two-particle Green-function in (\ref{BSeq}) is
\begin{eqnarray}
    G(p ; P) &=&  \frac{i}{(2\pi)^{4}}
 \left[ \frac{ \Pi^{(s_{a})}(\frac{1}{2} P+p)}
 { ( \frac{1}{2} P + p)^{2} - M_{a}^{2}+i\delta} \right]^{(a)}\cdot \nonumber\\ 
 && \times \left[ \frac{ \Pi^{(s_{b})}(\frac{1}{2} P-p)}
 { ( \frac{1}{2} P - p)^{2} - M_{b}^{2}+i\delta} \right]^{(b)}\ .
\label{Green} \end{eqnarray}
 
Using now a complete set of on-mass-shell spin s-states in the first
line of (\ref{Greens}) one finds that the Feynman propagator of a
spin-s baryon off-mass-shell can be written as \cite{BD65}
\begin{eqnarray}
   \frac{ \Pi^{(s)}(p) }{p^{2} -M^{2}+ i \delta} &=&
   \frac{M}{E({\bf p})}
 \left[\frac{\Lambda_{+}^{(s)}({\bf p})}{p_{0}-E({\bf p})+i\delta} 
 \right.\nonumber\\ &&\left. \hspace{1cm} 
 -\frac{\Lambda_{-}^{(s)}(-{\bf p})}{p_{0}+E({\bf p})-i\delta}\right]\ ,
\label{eq:20.7} \end{eqnarray}
 for $s= \frac{1}{2}, \frac{3}{2}, \ldots $.
Here, $\Lambda_{+}^{(s)}({\bf p})$ and $\Lambda_{-}^{(s)}({\bf p})$ are the
on-mass-shell projection operators on the positive- and
negative-energy states. For the nucleon 
they are 
\begin{eqnarray}
    \Lambda_{+}({\bf p}) &=& \sum_{\sigma =-1/2}^{+1/2}
    u({\bf p},\sigma)\otimes\bar{u}({\bf p},\sigma)\ , \nonumber\\
    \Lambda_{-}({\bf p}) &=& -\sum_{\sigma =-1/2}^{+1/2}
    v({\bf p},\sigma)\otimes\bar{v}({\bf p},\sigma)\ .                      
\label{projop1} \end{eqnarray} 
where $u({\bf p},\sigma)$ and $v({\bf p},\sigma)$ are the Dirac spinors for
spin-$\frac{1}{2}$ particles, and $E({\bf p})=\sqrt{{\bf p}^2+M^2}$
with $M$ the nucleon mass.
Then, in the cm-system, where ${\bf P}=0$ and $P_{0}=W$,
the Green-function can be written as

 \begin{widetext}

\begin{eqnarray}
 G(p;W) &=& \frac{i}{(2\pi)^4}
  \left(\frac{M_a}{E_a({\bf p})} \right)
   \left[ \frac{\Lambda_+^{(s_a)}({\bf p})}
{\frac{1}{2} W+p_0-E_a({\bf p})+i\delta}
    -\frac{\Lambda_-^{(s_a)}(-{\bf p})}
    {\frac{1}{2} W+p_0+E_a({\bf p})-i\delta}\right] \nonumber \\[0.4cm]
& & \hspace*{0.7cm} \times
  \left(\frac{M_b}{E_b({\bf p})}\right)
 \left[ \frac{\Lambda_+^{(s_b)}(-{\bf p})}
 {\frac{1}{2} W-p_0-E_b({\bf p})+i\delta}
    - \frac{\Lambda_-^{(s_b)}( {\bf p})}
    {\frac{1}{2} W-p_0+E_b({\bf p})-i\delta}\right]
 \label{prop1} \end{eqnarray} 

 \end{widetext}

Multiplying out (\ref{prop1})
we  write the ensuing terms in shorthand notation
\begin{eqnarray}
    G(p;W) &=& G_{++}(p;W) + G_{+-}(p;W) \nonumber\\
 && + G_{-+}(p;W) + G_{--}(p;W)\ ,
\label{eq:20.10} \end{eqnarray}
where $G_{++}$ etc. corresponds to the term with
$\Lambda_{+}^{s_{a}} \Lambda_{+}^{s_{b}}$ etc.
Introducing the wave-functions (see \cite{Salp52})
\begin{equation}
 \psi_{rs}(p')= \Lambda_{r}^{s_{a}} \Lambda_{s}^{s_{b}} \psi(p')
 \hspace{0.4cm}  (r,s=+,-) \ ,
\label{eq:20.11} \end{equation}
the two-body equation, (\ref{BSeq}) can be written for {\it e.g.}
 $\psi_{+ +}$ as
\begin{eqnarray}
    \psi_{++}(p)&=& \psi^{0}_{++}(p) + G_{++}(p;W)\cdot \nonumber\\
  && \times \int\! d^{4}p' \sum_{r,s}
  I(p,p')_{++,rs} \psi_{rs}(p')\ , 
\label{psiBS} \end{eqnarray}
and similar equations for $\psi_{+-},\ \psi_{-+}$, and $\psi_{--}$.
 
Invoking `dynamical pair-suppression', as discussed in
\cite{Rij91}, (\ref{psiBS}) reduces to a four-dimensional equation
for $\psi_{++}$, {\it i.e.}
\begin{eqnarray}
    \psi_{++}(p') &=& \psi^{0}_{++}(p') + G_{++}(p';W) \nonumber\\
 && \times \int d^{4}p\, I(p',p)_{++,++} \psi_{++}(p)\ ,
\label{eq:20.13} \end{eqnarray}
with the Green-function
\begin{eqnarray}
 G_{++}(p;W)&=& \frac{i}{(2\pi)^{4}}
 \left[\frac{M_{a}M_{b}}{E_{a}({\bf p})E_{b}({\bf p})}\right]
 \Lambda_{+}^{(s_{a})}({\bf p} ) \Lambda_{+}^{(s_{b})}(-{\bf p})
  \cdot   \nonumber \\[0.0cm]
 &\times&  \left[\frac{1}{2} W+p_{0}-E_{a}({\bf p})+i\delta\right]^{-1} \nonumber\\
 &\times&  \left[\frac{1}{2} W-p_{0}-E_{b}({\bf p})+i\delta\right]^{-1}\ .
 \label{Greenth} \end{eqnarray}
 
\subsection{Three-Dimensional Equation}
Following the same procedures as in \cite{Rij91} we introduce
the three-dimensional wave-function according to Salpeter \cite{Salp52}
by
\begin{equation}
  \phi({\bf p})= \sqrt{
  \frac{E_{a}({\bf p})E_{b}({\bf p})}{M_{a}M_{b}}}\,
  \int_{-\infty}^{ \infty} \psi(p_{\mu}) dp_{0} \ .
\label{eq:20.15} \end{equation}
Next, we use for the right-inverse of the $\int dp_0$-operation the {\it ansatz} 
proposed by Klein and Macke \cite{Klein53}, 
\begin{equation}
    \psi(p'_{\mu})=
    \sqrt{\frac{M_{a}M_{b}}{E_{a}({\bf p}')E_{b}({\bf p}')}}\,
    A_{W}(p'_{\mu}) \phi({\bf p}')\ ,
\label{Kansatz} \end{equation}
where the right-inverse is given by
\begin{eqnarray}
    A_{W}(p'_{\mu}) 
  &=& -\frac{1}{2\pi i} \frac{W-{\cal W}({\bf p}')}            
  {F^{(a)}_{W}({\bf p}',p_{0}') F^{(b)}_{W}(-{\bf p}',-p_{0}')}\ , 
\label{ansatz} \end{eqnarray}
with the frequently used notations           
\begin{eqnarray}
  &&  F_{W}({\bf p},p_{0}) = p_{0}-E({\bf p})+ \frac{1}{2} W + i\delta
 \nonumber\\
  &&  {\cal W}({\bf p}) = E_{a}({\bf p}) + E_{b}({\bf p})\ .
\label{eq:20.18} \end{eqnarray}
Applying now the $p_{0}$ integration to (\ref{eq:20.13}), and performing the 
$p'_{0}$-integration, made possible by the {\it ansatz} above, one arrives
at the Thompson equation \cite{Thompson70}
\begin{eqnarray}
    \phi_{++}({\bf p}') &=& \phi^{(0)}_{++}({\bf p}') +
    E_{2}^{(+)}({\bf p}'; W) \nonumber\\
 && \times \int \frac{d^{3}p}{(2\pi)^3}\; 
    K^{{\it irr}}({\bf p}',{\bf p}| W)\; \phi_{++}({\bf p})\ ,
\label{Thomp1} \end{eqnarray}
 where the Green function is

\begin{equation}
    E_{2}^{(+)}({\bf p}'; W)=  
 \frac{1}{(2\pi)^{3}}\;
    \frac{\Lambda_{+}^{a}( {\bf p}') \Lambda_{+}^{b}(-{\bf p}')}
    {\left(W-{\cal W}({\bf p}') +i\delta \right)}\ ,
 \label{Thompgreen} \end{equation}
and the kernel is given by

\begin{widetext}
 
\begin{eqnarray}
  K^{{\it irr}}({\bf p}',{\bf p}| W)&=& -\frac{1}{(2\pi)^{2}}
 \sqrt{\frac{M_{a}M_{b}}{E_{a}({\bf p}') E_{b}({\bf p}')} }
 \sqrt{\frac{M_{a}M_{b}}{E_{a}({\bf p}) E_{b}({\bf p})} }
 \left(W-{\cal W}({\bf p}')\right)\left(W-{\cal W}({\bf p})\right)
 \nonumber \\[0.2cm] &\times&
  \int_{-\infty}^{+\infty} dp_{0}'
   \int_{-\infty}^{+\infty} dp_{0} \left[ \vphantom{\frac{A}{A}}
 \left\{F_{W}^{(a)}({\bf p}',p_{0}')
 F_{W}^{(b)}(-{\bf p}',-p_{0}')\right\}^{-1}
 \right.\nonumber \\[0.2cm] &\times& \left.
 \left[ I(p_{0}',{\bf p}'; p_{0},{\bf p}) \right]_{++,++}
    \left\{F_{W}^{(a)}({\bf p},p_{0})
    F_{W}^{(b)}(-{\bf p},-p_{0})\right\}^{-1}
\vphantom{\frac{A}{A}}\right]\ .
\label{Thomp2}  \end{eqnarray}

\end{widetext}

 The $M/E$-factors in (\ref{Thomp2}) are due to the difference
between the relativistic and the non-relativistic normalization of
the two-particle states. In the following we simply put
$M/E({\bf p})=1$ in the kernel $K^{irr}$ Eq.~(\ref{Thomp2}). The corrections
to this approximation would give $(1/M)^{2}$-corrections
to the potentials, which we neglect in this paper. In the same approximation
there is no difference between the Thompson (\ref{Thomp1})
and the Lippmann-Schwinger equation, when the connection between these 
equations is made using multiplication factors. Henceforth, we will not
distinguish between the two.
 
The contributions to the two-particle irreducible kernel
$K^{{\it irr}}$ up to second order in the meson-exchange
are given in detail in \cite{RS96a}.
For the definition of the TME-potential in the
Lippmann-Schwinger equation we shall need the complete fourth-order
kernel for the Thompson equation (\ref{Thomp1}). In operator notation,
we have from (\ref{Thomp1})
\begin{eqnarray}
  \phi_{++}&=&\phi^{(0)}_{++} + E_{2}^{(+)}\; K^{{\it irr}}\; \phi_{++}
 \nonumber \\[0.2cm]
           &=&\phi^{(0)}_{++} + E_{2}^{(+)} \left(
  K^{{\it irr}} +  K^{{\it irr}}\; E_{2}^{(+)}\; K^{{\it irr}} + ....
  \right) \phi_{++}^{(0)}
 \nonumber \\[0.2cm]
           &\equiv& \left( 1 + E_{2}^{(+)}\; M \right) \phi_{++}^{(0)}
\label{Thomp3} \end{eqnarray} 
which implies for the complete kernel $M$ the integral equation
\begin{eqnarray}
    M({\bf p}',{\bf p}| W) &=& K^{{\it irr}}({\bf p}',{\bf p}| W) +
    \int \frac{d^{3}p''}{(2\pi)^3}\; K^{{\it irr}}({\bf p}',{\bf p}''| W)\cdot
 \nonumber\\ 
 && \times E_{2}^{(+)}({\bf p}''; W)\; M({\bf p}'',{\bf p}| W)
\label{Thompkern} \end{eqnarray}
 
 
\subsection{Lippmann-Schwinger Equation}

 \begin{figure}   
 \resizebox{5cm}{11.25cm}
 {\includegraphics[200,200][400,650]{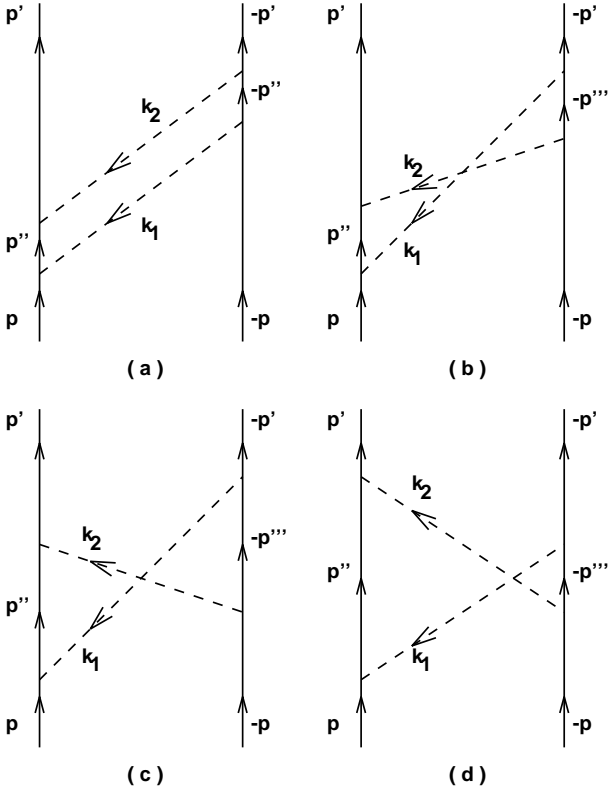}}
\caption{BW two-meson-exchange graphs: (a) planar and (b)--(d) crossed
        box. The dashed line with momentum ${\bf k}_{1}$ refers to the
        pion and the dashed line with momentum ${\bf k}_{2}$ refers
        to one of the other (vector, scalar, or pseudoscalar) mesons.
        To these we have to add the ``mirror'' graphs, and the
        graphs where we interchange the two meson lines.}
\label{bwfig}
 \end{figure}

 \begin{figure}   
 \resizebox{5cm}{5.75cm} 
 {\includegraphics[200,420][400,650]{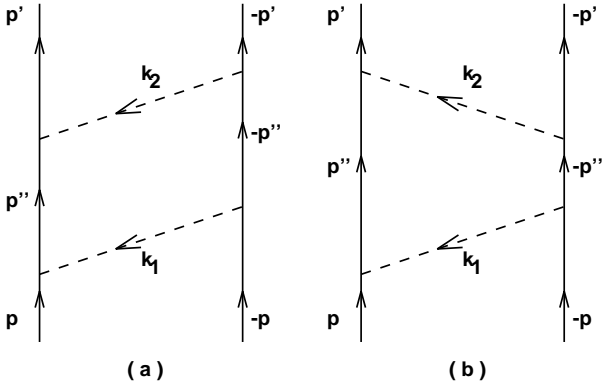}}
\caption{Planar-box TMO two-meson-exchange graphs.
         Same notation as in Fig.~\protect\ref{bwfig}.
         To these we have to add the ``mirror'' graphs, and the
         graphs where we interchange the two meson lines.}
 \label{tmofig}
 \end{figure}

The transformation of (\ref{Thompkern}) to the Lippmann-Schwinger 
equation can be effectuated by defining
\begin{eqnarray}
 T({\bf p}',{\bf p}) &=& N({\bf p}')\ M({\bf p}',{\bf p};W)\ N({\bf p})\ , 
 \nonumber\\[0.2cm]
 V({\bf p}',{\bf p}) &=& N({\bf p}')\ K^{irr}({\bf p}',{\bf p};W)\ N({\bf p})\ ,       
\label{eq:20.24} \end{eqnarray}
where the transformation function is 
\begin{equation} 
 N({\bf p}) = \sqrt{\frac{{\bf p}_i^2-{\bf p}^2}{2M_N(E\left({\bf p}_i)-E({\bf p})\right)}}\ .
\label{eq:20.25} \end{equation}
Application of this transformation, yields the Lippmann-Schwinger
equation
\begin{eqnarray}
    T({\bf p}',{\bf p}) &=& V({\bf p}',{\bf p}) +
    \int \frac{d^{3}p''}{(2\pi)^3} \nonumber\\ 
 && \times V({\bf p}',{\bf p}'')\ g({\bf p}''; W)\; T({\bf p}'',{\bf p})
\label{eq:20.26} \end{eqnarray}
with the standard Green function     
\begin{equation}
    g({\bf p};W) = \frac{1}{(2\pi)^{3}} \;
     \Lambda_{+}^{a}({\bf p}) \Lambda_{+}^{b}(-{\bf p})\
    \frac{M_N}{{\bf p}_i^{2}-{\bf p}^{2}+i\delta} \ .
\label{eq:20.27} \end{equation}
The corrections to the approximation $E_{2}^{(+)} \approx g({\bf p}; W)$ 
are of order $1/M^{2}$, which we neglect hencforth.
 
The transition from Dirac-spinors to
Pauli-spinors, is given in Appendix C of \cite{Rij91}, where we write for the 
the Lippmann-Schwinger equation in the 4-dimensional Pauli-spinor space
\begin{eqnarray}
 {\cal T}({\bf p}',{\bf p})&=&{\cal V}({\bf p}',{\bf p}) + \int \frac{d^{3} p''}{(2\pi)^3}\
 \nonumber\\ && \times 
 {\cal V}({\bf p}',{\bf p}'')\  \tilde{g}({\bf p}''; W)\
  {\cal T}({\bf p}'',{\bf p})\ .
\label{LSeq2} \end{eqnarray}

The ${\cal T}$-operator in Pauli spinor-space is defined by
\begin{eqnarray}
 && \chi^{(a)\dagger}_{\sigma_{a}}\chi^{(b)\dagger}_{\sigma_{b}}\; 
 {\cal T}({\bf p}',{\bf p})\;
 \chi^{(a)}_{\sigma'_{a}}\chi^{(b)}_{\sigma'_{b}}  =              
 \bar{u}_{a}({\bf p},\sigma_{a})\bar{u}_{b}(-{\bf p},\sigma_{b}) \nonumber\\
 && \times T({\bf p},{\bf p}'')\;
     u_{a}({\bf p}'',\sigma'_{a}) u_{b}(-{\bf p}'',\sigma'_{b}) \ .
 \label{LSitems} \end{eqnarray}
and similarly for the ${\cal V}$-operator.
Like in the derivation of the OBE-potentials \cite{NRS78,NRS77}
we make off-shell and on-shell the approximation,
  $ E({\bf p})= M + {\bf p}^{2}/2M $
 and $ W = 2\sqrt{{\bf p}_i^{2}+M^{2}} = 2M + {\bf p}_i^{2}/M$ ,     
everywhere in the interaction kernels, which, of course,
is fully justified for low energies only. 
In contrast to these kind of approximations, of course the full
${\bf k}^{2}$-dependence of the form factors is kept
throughout the derivation of the TME. 
Notice that the Gaussian form factors suppress the high momentum
transfers strongly. This means that the contribution to the potentials
from intermediate states which are far off-energy-shell can not
be very large. 

Because of rotational invariance and parity conservation, the $T$-matrix, which is
a $4\times 4$-matrix in Pauli-spinor space, can be expanded 
into the following set of in general 8 spinor invariants, see for example 
\cite{SNRV71,MRS89}. Introducing \cite{notation1}
\begin{equation}
  {\bf q}=\frac{1}{2}({\bf p}_{f}+{\bf p}_{i})\ , \
  {\bf k}={\bf p}_{f}-{\bf p}_{i}\ , \           
  {\bf n}={\bf p}_{i}\times {\bf p}_{f}\ ,
\label{eq:20.30} \end{equation}
with, of course, ${\bf n}={\bf q}\times {\bf k}$,
we choose for the operators $P_{j}$ in spin-space
\begin{equation}
\begin{array}{ll}
  P_{1}=1  &  P_{2}= 
 \mbox{\boldmath $\sigma$}_1\cdot\mbox{\boldmath $\sigma$}_2 \\[0.2cm]
 P_{3}=(\mbox{\boldmath $\sigma$}_1\cdot{\bf k})(\mbox{\boldmath $\sigma$}_2\cdot{\bf k})
 -\frac{1}{3}(\mbox{\boldmath $\sigma$}_1\cdot\mbox{\boldmath $\sigma$}_2) 
  {\bf k}^2 & P_{4}=\frac{i}{2}(\mbox{\boldmath $\sigma$}_1+
 \mbox{\boldmath $\sigma$}_2)\cdot{\bf n} \\[0.2cm]
 P_{5}=(\mbox{\boldmath $\sigma$}_1\cdot{\bf n})(\mbox{\boldmath $\sigma$}_2\cdot{\bf n})
  & P_{6}=\frac{i}{2}(\mbox{\boldmath $\sigma$}_1-
 \mbox{\boldmath $\sigma$}_2)\cdot{\bf n} \\[0.2cm]
 \multicolumn{2}{l}{
 P_{7}=(\mbox{\boldmath $\sigma$}_1\cdot{\bf q})(\mbox{\boldmath $\sigma$}_2\cdot{\bf k})
 +(\mbox{\boldmath $\sigma$}_1\cdot{\bf k})(\mbox{\boldmath $\sigma$}_2\cdot{\bf q})
 } \\[0.2cm]
 \multicolumn{2}{l}{
 P_{8}=(\mbox{\boldmath $\sigma$}_1\cdot{\bf q})(\mbox{\boldmath $\sigma$}_2\cdot{\bf k})
 -(\mbox{\boldmath $\sigma$}_1\cdot{\bf k})(\mbox{\boldmath $\sigma$}_2\cdot{\bf q})
 }  
\end{array}
\label{eq:20.31} \end{equation}
Here we follow \cite{MRS89}, where in contrast to \cite{NRS78},
we have chosen $P_{3}$ to be a purely `tensor-force' operator.
The expansion in spinor-invariants reads
\begin{equation}
 {\cal T}({\bf p}_f,{\bf p}_i) = \sum_{j=1}^8\ \widetilde{T}_j({\bf p}_f^2,{\bf p}_i^2,
 {\bf p}_f\cdot{\bf p}_i)\ P_j({\bf p}_f,{\bf p}_i)\ .
\label{eq:20.32} \end{equation}
Similarly to (\ref{eq:20.32}) we expand the potentials $V$. Again
following \cite{MRS89}, we neglect the potential forms $P_{7}$ and
$P_{8}$, and also the dependence of the potentials on
${\bf k}\cdot{\bf q}$ . Then, the expansion (\ref{eq:20.32}) reads for
the potentials as follows
\begin{equation}
  {\cal V} =\sum_{j=1}^{4}\tilde{V}_{j}({\bf k}\,^{2},{\bf q}\,^{2})\
 P_{j}({\bf k},{\bf q})\ .
\label{eq:20.33} \end{equation}
 
We develop in the following subsections a new representation of the
TME-potentials.
Included are the BW-graphs shown in Fig.~\ref{bwfig}, 
and the TMO-graphs shown in Fig.~\ref{tmofig}. 
Also the employed notations for the momenta are indicated in the figures.

 \begin{widetext}

\section{ Momentum Space Representation TME-potentials}        
In this section  we give an outline of the procedures in making
the partial wave analysis of TME-potentials in momentum space.

\subsection{ TPS Central Potentials}        
\label{app:B3}
Consider the basic integral
\begin{eqnarray}
\tilde{V}({\bf k}) &=& \int\int\frac{d^3k_1d^3k_2}{(2\pi)^3}\
 \delta({\bf k - k_1 - k_2})\ 
 \tilde{F}({\bf k}_1^2) \tilde{G}({\bf k}_2^2) 
 = \int \frac{d^3 \Delta}{(2\pi)^3}\
 \tilde{F}\left(\mbox{\boldmath $\Delta$}^2\right)\cdot
 \tilde{G}\left(({\bf k}-\mbox{\boldmath $\Delta$})^2\right)\ ,           
\label{AppB.16a} \end{eqnarray}
with $\tilde{F}({\bf k}_1)$ and $\tilde{G}({\bf k}_2)$ having the                    
generic soft-core forms 
\begin{equation}
 \tilde{F}({\bf k}^2) = \frac{\exp\left[-{\bf k}^2/\Lambda_1^2\right]}{{\bf k}^2+m_1^2}\ ,\
 \tilde{G}({\bf k}^2) = \frac{\exp\left[-{\bf k}^2/\Lambda_2^2\right]}{{\bf k}^2+m_2^2}\ .  
\label{AppB.16b} \end{equation}
Exploiting the identity
\begin{equation}
 \frac{\exp\left[-{\bf k}^2/\Lambda^2\right]}{{\bf k}^2+m^2} = 
 e^{m^2/\Lambda^2} \int_1^\infty \frac{dt}{\Lambda^2}\ 
 \exp\left[-\left(\frac{{\bf k}^2+m^2}{\Lambda^2}\right) t\right]
\label{AppB.17} \end{equation}
one can write (\ref{AppB.16a}) as
\begin{eqnarray}
 \tilde{V}({\bf k}) &=& e^{m_1^2/\Lambda_1^2} e^{m_2^2/\Lambda_2^2} 
 \int_1^\infty \frac{dt}{\Lambda_1^2} \int_1^\infty \frac{du}{\Lambda_2^2}\
 e^{-(m_1^2/\Lambda_1^2)t} e^{-(m_2^2/\Lambda_2^2)u}\cdot \nonumber\\
 && \times \int \frac{d^3 \Delta}{(2\pi)^3}\
 \exp\left[-\left(\frac{t}{\Lambda_1^2}+\frac{u}{\Lambda_2^2}\right)
 \mbox{\boldmath $\Delta$}^2 + \left(2{\bf k}\cdot\mbox{\boldmath $\Delta$}-{\bf k}^2\right)
 \frac{u}{\Lambda_2^2}\right]\ .          
\label{AppB.18} \end{eqnarray}
 The $\Delta$-integral in (\ref{AppB.18}) has been evaluated above, see (\ref{AppB.12}), and is 
\begin{eqnarray*}
 && (4\pi a)^{-3/2}\ \exp\left[-\frac{t u/\Lambda_1^2\Lambda_2^2}
 {\left(t/\Lambda_1^2+u/\Lambda_2^2\right)}{\bf k}^2\right]\ , \       
  {\rm with}\ a = \frac{t}{\Lambda_1^2} + \frac{u}{\Lambda_2^2}\ .
\end{eqnarray*}
Redefining the variables $t\rightarrow t/\Lambda_1^2$ and 
$u\rightarrow u/\Lambda_2^2$, we rewrite (\ref{AppB.18}) in the form
\begin{eqnarray}
 \tilde{V}({\bf k}) &=& (4\pi)^{-3/2}\ e^{m_1^2/\Lambda_1^2} e^{m_2^2/\Lambda_2^2} 
 \int_{t_0}^\infty dt \int_{u_0}^\infty du\
 \frac{\exp[-(m_1^2t+m_2^2 u)]}{(t+u)^{3/2}}
\cdot \nonumber\\ && \times 
 \exp\left[-\left(\frac{tu}{t+u}\right){\bf k}^2\right]\ \  \ (t_0=1/\Lambda_1^2,\
 u_0=1/\Lambda_2^2)\ .
\label{AppB.19} \end{eqnarray}
{\it This form we consider as the basic representation of the soft-core TME-potentials
in momentum space}.

\subsection{ TPS Tensor and Spin-spin Potentials}        
\label{app:B4}
In this subsection we apply the same technique as used in the previous one
to the more complicated case involving the tensor interaction. 
For ps-ps in momentum space we have
\begin{eqnarray}
\tilde{V}({\bf k}) &=& \int\int\frac{d^3k_1d^3k_2}{(2\pi)^3}\
 \delta({\bf k - k_1 - k_2})\ \tilde{F}({\bf k}_1^2) \tilde{G}({\bf k}_2^2)\cdot
 \left[\mbox{\boldmath $\sigma$}_1\cdot{\bf k}_1\times{\bf k}_2\right] 
 \left[\mbox{\boldmath $\sigma$}_2\cdot{\bf k}_1\times{\bf k}_2\right] 
 \nonumber\\ &=& \int \frac{d^3 \Delta}{(2\pi)^3}\
 \left[\mbox{\boldmath $\sigma$}_1\cdot\mbox{\boldmath $\Delta$}\times{\bf k}\right]
 \left[\mbox{\boldmath $\sigma$}_2\cdot\mbox{\boldmath $\Delta$}\times{\bf k}\right]
 \tilde{F}\left(\mbox{\boldmath $\Delta$}^2\right)\cdot
 \tilde{G}\left( ({\bf k}-\mbox{\boldmath $\Delta$})^2\right)\ ,           
\label{AppB.20} \end{eqnarray}
which leads to the basic integral to be evaluated:
\begin{eqnarray}
\tilde{V}_{mn}({\bf k}) &=& \int \frac{d^3 \Delta}{(2\pi)^3}\
 \Delta_m \Delta_n\ \tilde{F}\left(\mbox{\boldmath $\Delta$}^2\right)\cdot
 \tilde{G}\left(({\bf k}-\mbox{\boldmath $\Delta$})^2\right)\ .           
\label{AppB.21} \end{eqnarray}
Repeating the steps taken in the previous subsection, one arrives at the 
analogue of (\ref{AppB.18}):
\begin{eqnarray}
\tilde{V}_{mn}({\bf k}) &=& e^{m_1^2/\Lambda_1^2} e^{m_2^2/\Lambda_2^2} 
 \int_1^\infty \frac{dt}{\Lambda_1^2} \int_1^\infty \frac{du}{\Lambda_2^2}\
 e^{-(m_1^2/\Lambda_1^2)t} e^{-(m_2^2/\Lambda_2^2)u}\cdot \nonumber\\
 && \times \int \frac{d^3 \Delta}{(2\pi)^3}\ \Delta_m \Delta_n\
 \exp\left[-\left(\frac{t}{\Lambda_1^2}+\frac{u}{\Lambda_2^2}\right)
 \mbox{\boldmath $\Delta$}^2 + \left(2{\bf k}\cdot\mbox{\boldmath $\Delta$}-{\bf k}^2\right)
 \frac{u}{\Lambda_2^2}\right]\ .          
\label{AppB.22} \end{eqnarray}
Using the integral
\begin{eqnarray*}
 && \int \frac{d^3 \Delta}{(2\pi)^3}\ \Delta_m \Delta_n\
 \exp\left[-a \mbox{\boldmath $\Delta$}^2 + 
 2b{\bf k}\cdot\mbox{\boldmath $\Delta$}\right] = 
  (4\pi a)^{-3/2}\ \frac{1}{a}
 \left[\frac{1}{2}\delta_{mn} + \frac{b^2}{a} k_m k_n \right] 
 \exp\left[+\frac{b^2}{a} {\bf k}^2\right]\  
 \end{eqnarray*}
one obtains, again after the redefinitions $t \rightarrow t/\Lambda_1^2$ and
 $u \rightarrow u/\Lambda_2^2$, for $\tilde{V}_{mn}$ the expression
\begin{eqnarray}
\tilde{V}_{mn}({\bf k}) &=& 
 (4\pi)^{-3/2}e^{m_1^2/\Lambda_1^2} e^{m_2^2/\Lambda_2^2} 
 \int_{t_0}^\infty dt \int_{u_0}^\infty du\ 
\frac{\exp[-(m_1^2t+m_2^2u)]}{(t+u)^{5/2}}
 \cdot\nonumber\\ && \times 
\left[\frac{1}{2}\delta_{mn}+\frac{u^2}{(t+u)} k_m k_n\right]\
 \exp\left[-\left(\frac{tu}{t+u}\right){\bf k}^2\right]\ .
\label{AppB.23} \end{eqnarray}
It is clear that in (\ref{AppB.20}) only the $\delta_{mn}$-term contributes. The result
is 
\begin{eqnarray}
 \tilde{V}({\bf k}) &=& 
 - \left[ \left(\mbox{\boldmath $\sigma$}_1\cdot{\bf k}\right)\
 \left(\mbox{\boldmath $\sigma$}_2\cdot{\bf k}\right) - \frac{1}{3}
 \left(\mbox{\boldmath $\sigma$}_1\cdot\mbox{\boldmath $\sigma$}_2\right)\
 {\bf k}^2\right]\  \tilde{H}({\bf k}^2) 
 +\frac{2}{3}
 \left(\mbox{\boldmath $\sigma$}_1\cdot\mbox{\boldmath $\sigma$}_2\right)\
 {\bf k}^2\ \tilde{H}({\bf k}^2) 
\label{AppB.24} \end{eqnarray}
with 
\begin{eqnarray}
\tilde{H}({\bf k}^2) &=& 
 \frac{1}{2}(4\pi)^{-3/2}e^{m_1^2/\Lambda_1^2} e^{m_2^2/\Lambda_2^2} 
 \int_{t_0}^\infty\; dt \int_{u_0}^\infty\; du\ 
\frac{\exp[-(m_1^2t+m_2^2u)]}{(t+u)^{5/2}}
 \cdot\nonumber\\ && \times 
 \exp\left[-\left(\frac{tu}{t+u}\right){\bf k}^2\right]\ .
\label{AppB.25} \end{eqnarray}
{}From equations (\ref{AppB.24})-(\ref{AppB.25}) one can read off immediately the
projection of (\ref{AppB.20}) onto the spinor invariants (\ref{eq:20.32}).
In Appendix~\ref{app:F} the same result is derived in an alternative way starting
from the configuration space potentials.

\section{ General PS-PS Exchange Potentials}        
\label{app:B5}
In \cite{RS96a}) the derivation of
the ps-ps exchange potentials both in momentum and in configuration space is given.       
In this reference the configuration space potentials are worked out fully.
The topic of this paper is to do the same for the momentum space description.
In particular, the partial wave analysis is performed leading to a 
representation which is very suitable for numerical evaluation.\\
{}From \cite{RS96a} it appears that the momentum space TME-potential can be 
represented in general in the form
\begin{eqnarray}
\widetilde{V}_{\alpha\beta}({\bf k}) &=& \frac{2}{\pi}\int_0^\infty d\lambda\ 
 \sum_{i=1}^N\ w_i (\lambda) \int \frac{d^3 \Delta}{(2\pi)^3}\ 
 \widetilde{O}_{\alpha\beta}({\bf k},\mbox{\boldmath $\Delta$})\ 
 \widetilde{F}_{\alpha}^{(i)}(\mbox{\boldmath $\Delta$}^2,\lambda)\
 \widetilde{F}_{\beta}^{(i)}(({\bf k}-\mbox{\boldmath $\Delta$})^2,\lambda)\ .
\label{AppB.26} \end{eqnarray}
Here, $\sum_{i}$ stands for the number of different types of $\lambda$-functions, and
$w_i(\lambda)$ are the corresponding weights. The operators $\widetilde{O}_{\alpha\beta}$
are for example in the case of ps-ps exchange given by, see \cite{RS96a} Table II,
\begin{eqnarray}
 \widetilde{O}_{\alpha\beta}^{//}({\bf k},\mbox{\boldmath $\Delta$}) &=&
 2\left( \mbox{\boldmath $\Delta$}\cdot({\bf k}-\mbox{\boldmath $\Delta$})\right)^2 -
 2\left[\mbox{\boldmath $\sigma$}_1\cdot\mbox{\boldmath $\Delta$}\times{\bf k}\right]
 \left[\mbox{\boldmath $\sigma$}_2\cdot\mbox{\boldmath $\Delta$}\times{\bf k}\right]\ ,
 \nonumber\\ 
 \widetilde{O}_{\alpha\beta}^{X}({\bf k},\mbox{\boldmath $\Delta$}) &=&
 2\left( \mbox{\boldmath $\Delta$}\cdot({\bf k}-\mbox{\boldmath $\Delta$})\right)^2 +
 2\left[\mbox{\boldmath $\sigma$}_1\cdot\mbox{\boldmath $\Delta$}\times{\bf k}\right]
 \left[\mbox{\boldmath $\sigma$}_2\cdot\mbox{\boldmath $\Delta$}\times{\bf k}\right]\ .
\label{AppB.27} \end{eqnarray}
In \cite{footnote} we remind the reader of the precise contents of these operators.
As an example for the $\sum_i$ we give explicitly the case of $\pi\eta$-exchange, the 
planar BW- and TMO-graphs, and the BW-crossed graphs. These can be written as
\cite{RS96a}
\begin{eqnarray}
\widetilde{V}_{\pi\eta}^{//,X}({\bf k}) &=& \pm\frac{2}{\pi}\int_0^\infty 
 \frac{d\lambda}{\lambda^2}\ 
 \int \frac{d^3 \Delta}{(2\pi)^3}\ 
 \widetilde{O}_{\pi\eta}^{//,X}({\bf k},\mbox{\boldmath $\Delta$})\     
\left[ \widetilde{F}_{\pi}(\mbox{\boldmath $\Delta$}^2,m_\pi)\  
 \widetilde{F}_{\eta}(({\bf k}-\mbox{\boldmath $\Delta$})^2,m_\eta)\   
 \right. \nonumber\\ && \left. -
 \widetilde{F}_{\pi}(\mbox{\boldmath $\Delta$}^2,\sqrt{m_\pi^2+\lambda^2})\   
 \widetilde{F}_{\eta}(({\bf k}-\mbox{\boldmath $\Delta$})^2,\sqrt{m_\eta^2+\lambda^2})\
 e^{-\lambda^2/\Lambda_\pi^2} e^{-\lambda^2/\Lambda_\eta^2}\right]\ .
\label{AppB.28} \end{eqnarray}
Here, the $(+)$-sign refers to the parallel graphs $(//)$, and the $(-)$-sign to the
crossed $(X)$ graphs.

\subsection{ Integration over $\lambda$-parameter}      
\label{app:B6}
Since, due to the use of (\ref{AppB.17}) and gaussian form factors all integrations
over $\lambda$ become gaussian, they can be performed analytically. Consider for example,
the type of integral appearing in (\ref{AppB.28}). We write
\begin{eqnarray}
\widetilde{V}_{\alpha\beta}({\bf k}) &=& \frac{2}{\pi}\int_0^\infty 
 \frac{d\lambda}{\lambda^2}\ 
 \int \frac{d^3 \Delta}{(2\pi)^3}\ 
\left[ \widetilde{F}_{\alpha}(\mbox{\boldmath $\Delta$}^2,m_1)\  
 \widetilde{F}_{\beta}(({\bf k}-\mbox{\boldmath $\Delta$})^2,m_2)\   
 \right. \nonumber\\ && \left. -
 \widetilde{F}_{\alpha}(\mbox{\boldmath $\Delta$}^2,\sqrt{m_1^2+\lambda^2})\   
 \widetilde{F}_{\beta}(({\bf k}-\mbox{\boldmath $\Delta$})^2,\sqrt{m_2^2+\lambda^2})\
 e^{-\lambda^2/\Lambda_1^2} e^{-\lambda^2/\Lambda_2^2}\right]\ .
\label{AppB.29} \end{eqnarray}
Going through the same steps as between equations (\ref{AppB.17})-(\ref{AppB.19}), we get 
for the second part between the square brackets, an extra $\lambda$-dependent factor
\[
  e^{-(\lambda^2/\Lambda_1^2)t} e^{-(\lambda^2/\Lambda_2^2)u} \rightarrow
  e^{-\lambda^2(t+u)}\ ,          
\]
where we applied again the same variable redefinition as before, i.e.            
$t \rightarrow t/\Lambda_1^2$ and $u \rightarrow u/\Lambda_2^2$. Then, 
the needed $\lambda$-integral in (\ref{AppB.29}) is
\begin{equation} 
\frac{2}{\pi}\int_0^\infty \frac{d\lambda}{\lambda^2}\left[1-
  e^{-(t+u)\lambda^2}\right] = \frac{2}{\sqrt{\pi}}\sqrt{t+u}\ .
\label{AppB.29b} \end{equation}
Therefore (\ref{AppB.29}) becomes, compare with (\ref{AppB.16a})-(\ref{AppB.19}),
\begin{eqnarray}
\tilde{V}({\bf k}) &=& (2\pi)^{-2}\ e^{m_1^2/\Lambda_1^2} e^{m_2^2/\Lambda_2^2} 
 \int_{t_0}^\infty dt \int_{u_0}^\infty du\
 \frac{\exp[-(m_1^2t+m_2^2 u)]}{(t+u)}
\cdot \nonumber\\ && \times
 \exp\left[-\left(\frac{tu}{t+u}\right){\bf k}^2\right]\ \  \ (t_0=1/\Lambda_1^2,\
 u_0=1/\Lambda_2^2)\ .
\label{AppB.30} \end{eqnarray}
Similarly, all $\lambda$-integrations can be performed explicitly.

\section{ Projection PS-PS Exchange on Spinor Invariants}
The different contributions, as adiabatic, non-adiabatic, pseudovector-vertex
corrections, and off-shell potentials, are in direct correspondence with their
configuration space analogs, given in \cite{RS96a}. 

\subsection{ Adiabatic PS-PS Exchange Potentials}        
\label{app:B7}
We now have prepared all the necessary ingredients for putting things together
and to list the projection of ps-ps-exchange onto the potentials $V_{j}$.
We write
\begin{eqnarray}
\tilde{V}_{j}^{(0)}({\bf k}) &=& 
(4\pi)^{-3/2}\ e^{m_1^2/\Lambda_1^2} e^{m_2^2/\Lambda_2^2} 
\int_{t_0}^\infty dt\ \int_{u_0}^\infty du\ 
\frac{\exp[-(m_1^2 t+ m_2^2 u)]}{(t+u)^{3/2}}\cdot \nonumber\\ 
 && \times \exp\left[-\left(\frac{tu}{t+u}\right){\bf k}^2\right]\
\Omega_{j}^{(0)}({\bf k}^2;t,u)\ ,
\label{AppB.31} \end{eqnarray}
where 
 
\begin{eqnarray}
\Omega_1^{(0,//)}({\bf k}^2;t,u) &=& \frac{2}{\sqrt{\pi}}\ C_{NN,//}^{(I)}
\left(\frac{f_{NN\pi}}{m_\pi}\right)^2 \left(\frac{f_{NN\eta}}{m_\pi}\right)^2\cdot
 \left\{ \frac{15}{4}\ + 
\right.\nonumber\\ &&\nonumber\\ && \left.
 +\frac{1}{2}\left(\frac{t^2-8ut+u^2}{t+u}\right){\bf k}^2 +\left(\frac{t^2u^2}{(t+u)^2}\right)
 {\bf k}^4\right\}\cdot\frac{\sqrt{t+u}}{(t+u)^2}\ ,
\nonumber\\ &&\nonumber\\ 
\Omega_2^{(0,//)}({\bf k}^2;t,u) &=& -\frac{2}{\sqrt{\pi}}\ C_{NN,//}^{(I)}
\left(\frac{f_{NN\pi}}{m_\pi}\right)^2 \left(\frac{f_{NN\eta}}{m_\pi}\right)^2
\cdot \frac{1}{3} {\bf k}^2\cdot\frac{\sqrt{t+u}}{(t+u)}\ ,
\nonumber\\ &&\nonumber\\
\Omega_3^{(0,//)}({\bf k}^2;t,u) &=& \frac{2}{\sqrt{\pi}}\ C_{NN,//}^{(I)}
\left(\frac{f_{NN\pi}}{m_\pi}\right)^2 \left(\frac{f_{NN\eta}}{m_\pi}\right)^2
\cdot \frac{1}{2}\frac{\sqrt{t+u}}{(t+u)}\ ,
\label{AppB.32a} \end{eqnarray}
where $C_{NN,//}(I)$ denotes the isospin factor.\\

Expressions similar to (\ref{AppB.32a})  hold for the contribution of the 
crossed BW-graphs (X), and are given below. Note the difference in the $\tilde{O}$-operators 
in (\ref{AppB.27}) and $D_X(\omega_1,\omega_2)=-D_{//}(\omega_1,\omega_2)$.
 
\begin{eqnarray}
\Omega_1^{(0,X)}({\bf k}^2;t,u) &=& -\frac{2}{\sqrt{\pi}}\ C_{NN,X}^{(I)}
\left(\frac{f_{NN\pi}}{m_\pi}\right)^2 \left(\frac{f_{NN\eta}}{m_\pi}\right)^2\cdot
 \left\{ \frac{15}{4}\ + 
\right.\nonumber\\ &&\nonumber\\ && \left.
 +\frac{1}{2}\left(\frac{t^2-8ut+u^2}{t+u}\right){\bf k}^2+\left(\frac{t^2u^2}{(t+u)^2}\right)
 {\bf k}^4\right\}\cdot\frac{\sqrt{t+u}}{(t+u)^2}\ ,
\nonumber\\ &&\nonumber\\ 
\Omega_2^{(0,X)}({\bf k}^2;t,u) &=& -\frac{2}{\sqrt{\pi}}\ C_{NN,X}^{(I)}
\left(\frac{f_{NN\pi}}{m_\pi}\right)^2 \left(\frac{f_{NN\eta}}{m_\pi}\right)^2
\cdot \frac{1}{3} {\bf k}^2\cdot\frac{\sqrt{t+u}}{(t+u)}\ ,
\nonumber\\ &&\nonumber\\
\Omega_3^{(0,X)}({\bf k}^2;t,u) &=& \frac{2}{\sqrt{\pi}}\ C_{NN,X}^{(I)}
\left(\frac{f_{NN\pi}}{m_\pi}\right)^2 \left(\frac{f_{NN\eta}}{m_\pi}\right)^2
\cdot \frac{1}{2}\frac{\sqrt{t+u}}{(t+u)}\ .
\label{AppB.32b} \end{eqnarray}

It is convenient to introduce the expansions
\begin{equation}
\Omega_j^{(//,X)}({\bf k}^2;t,u) = \frac{2}{\sqrt{\pi}} C_{NN}^{(I)}(//,X)
\left(\frac{f_{NN\pi}}{m_\pi}\right)^2 \left(\frac{f_{NN\eta}}{m_\pi}\right)^2\cdot
 \sum_{k=0}^K\ \Upsilon_{j,k}^{(0),1a,1b,1c}(t,u)\ \left({\bf k}^2\right)^k\ .
\label{AppB.32c} \end{equation}
The functions $\Upsilon_{j,k}(t,u)$ are given in Table~\ref{table1}-\ref{table2}.

\subsection{ Non-adiabatic corrections}                       
\label{app:B8}
The denominators $D^{(1)}(\omega_1,\omega_2)$ for the non-adiabatic corrections
are given in Ref.~\cite{RS96a}, Table III. In appendix B of \cite{Rij91} 
it is explained how $1/\omega^4$ has to be treated. It follows that              
\begin{equation}
 D^{(1)}_{//}(\omega_1,\omega_2) \Rightarrow -\frac{1}{2}
 \left(\frac{\partial}{\partial m_1^2}-\frac{1}{\Lambda_1^2} +
 \frac{\partial}{\partial m_2^2}-\frac{1}{\Lambda_2^2}\right) 
 \frac{1}{\omega_1^2\omega_2^2}
\label{AppB.33a} \end{equation}
and one finds from (\ref{AppB.18}) and (\ref{AppB.33a}) 
that the contributions to the $\tilde{V}^{(1)}_j$ 
can be written in the form
\begin{eqnarray}
&& \tilde{V}_{j}^{(1a)}({\bf k}) = 
 (4\pi)^{-3/2}\ e^{m_1^2/\Lambda_1^2} e^{m_2^2/\Lambda_2^2} 
\int_{t_0}^\infty dt\ \int_{u_0}^\infty du\ 
\frac{\exp[-(m_1^2 t+ m_2^2 u)]}{(t+u)^{3/2}} \cdot  
 \nonumber\\ && \nonumber\\ && \times \frac{1}{2}\left(t+u\right)\cdot
 \exp\left[-\left(\frac{tu}{t+u}\right){\bf k}^2\right]\
\Omega_{j}^{(1a)}({\bf k}^2;t,u)\ .
\label{AppB.33} \end{eqnarray}
 The $\Omega_j^{(1a)}$ are given by
\begin{eqnarray}
&& \Omega_{1}^{(1a,//)}({\bf k}^2;t,u) = C_{NN,//}^{(I)}
\left(\frac{f_{NN\pi}}{m_\pi}\right)^2 \left(\frac{f_{NN\eta}}{m_\pi}\right)^2
\left(\frac{1}{M}\right)\cdot \left\{ -\frac{105}{8}\ + 
 \right.\nonumber\\ &&\nonumber\\ && \left.
 -\frac{15}{4}\left(\frac{t^2-5ut+u^2}{t+u}\right){\bf k}^2
 +\frac{3}{2}tu\left(\frac{t^2-5ut+u^2}{(t+u)^2}\right){\bf k}^4
 +\left(\frac{t^3u^3}{(t+u)^3}\right) {\bf k}^6\right\}\cdot\frac{1}{(t+u)^3}\ ,
\nonumber\\ &&\nonumber\\ 
&& \Omega_{2}^{(1a,//)}({\bf k}^2;t,u) = C_{NN,//}^{(I)}
\left(\frac{f_{NN\pi}}{m_\pi}\right)^2 \left(\frac{f_{NN\eta}}{m_\pi}\right)^2
\left(\frac{1}{M}\right)
\cdot -\frac{2}{3} {\bf k}^2\left\{-\frac{5}{4}+\frac{1}{2}\frac{tu}{t+u}{\bf k}^2\right\}
\cdot\frac{1}{(t+u)^2}\ ,
\nonumber\\ &&\nonumber\\
&& \Omega_{3}^{(1a,//)}({\bf k}^2;t,u) = C_{NN,//}^{(I)}
\left(\frac{f_{NN\pi}}{m_\pi}\right)^2 \left(\frac{f_{NN\eta}}{m_\pi}\right)^2
\left(\frac{1}{M}\right)
\cdot -\left\{\frac{5}{4}-\frac{1}{2}\frac{tu}{t+u}{\bf k}^2\right\}
\cdot\frac{1}{(t+u)^2}\ ,
\nonumber\\ &&\nonumber\\
&& \Omega_{4}^{(1a,//)}({\bf k}^2;t,u) = C_{NN,//}^{(I)}
\left(\frac{f_{NN\pi}}{m_\pi}\right)^2 \left(\frac{f_{NN\eta}}{m_\pi}\right)^2
\left(\frac{1}{M}\right)
\cdot -\left\{5-2\frac{tu}{t+u}{\bf k}^2\right\}
\cdot\frac{1}{(t+u)^2}\ .
\label{AppB.34a} \end{eqnarray}
Again, similar expressions hold for the contribution of the crossed BW-graphs (X), and are
given below. See for the differences in the $\tilde{O}^{(1)}$-operators \cite{RS96a} 
equations (5.3) and (5.4), 
and $D_X^{(1)}(\omega_1,\omega_2)=-2D_{//}^{(1)}(\omega_1,\omega_2)$,
cfrm. Table III of \cite{RS96a}.
\begin{eqnarray}
&& \Omega_{1}^{(1a,X)}({\bf k}^2;t,u) = -2C_{NN,X}^{(I)}
\left(\frac{f_{NN\pi}}{m_\pi}\right)^2 \left(\frac{f_{NN\eta}}{m_\pi}\right)^2
\left(\frac{1}{M}\right)\cdot \left\{ -\frac{105}{8}\ + 
 \right.\nonumber\\ &&\nonumber\\ && \left.
 -\frac{15}{4}\left(\frac{t^2-5ut+u^2}{t+u}\right){\bf k}^2
 +\frac{3}{2}tu\left(\frac{t^2-5ut+u^2}{(t+u)^2}\right){\bf k}^4
 +\left(\frac{t^3u^3}{(t+u)^3}\right) {\bf k}^6\right\}\cdot\frac{1}{(t+u)^3}\ ,
\nonumber\\ &&\nonumber\\ 
&& \Omega_{2}^{(1a,X)}({\bf k}^2;t,u) = +2C_{NN,X}^{(I)}
\left(\frac{f_{NN\pi}}{m_\pi}\right)^2 \left(\frac{f_{NN\eta}}{m_\pi}\right)^2
\left(\frac{1}{M}\right)
\cdot -\frac{2}{3} {\bf k}^2\left\{-\frac{5}{4}+\frac{1}{2}\frac{tu}{t+u}{\bf k}^2\right\}
\cdot\frac{1}{(t+u)^2}\ ,
\nonumber\\ &&\nonumber\\
&& \Omega_{3}^{(1a<X)}({\bf k}^2;t,u) = +2C_{NN,X}^{(I)}
\left(\frac{f_{NN\pi}}{m_\pi}\right)^2 \left(\frac{f_{NN\eta}}{m_\pi}\right)^2
\left(\frac{1}{M}\right)
\cdot -\left\{\frac{5}{4}-\frac{1}{2}\frac{tu}{t+u}{\bf k}^2\right\}
\cdot\frac{1}{(t+u)^2}\ .
\label{AppB.34b} \end{eqnarray}

\subsection{ Pseudovector-vertex corrections}                 
\label{app:B9}
Here, only the crossed BW-graphs contribute \cite{RS96a}.
{}From the denominators $D^{(1)}(\omega_1,\omega_2)=1/\omega_1^2\omega_2^2$
for the pseudovector-vertex corrections \cite{RS96a} 
one finds that the contributions to the $\tilde{V}^{(1)}_j$ 
can be written in the form
\begin{eqnarray}
\tilde{V}_{j}^{(1b)}({\bf k}) &=& 
(4\pi)^{-3/2}\ e^{m_1^2/\Lambda_1^2} e^{m_2^2/\Lambda_2^2}  
\int_{t_0}^\infty dt\ \int_{u_0}^\infty du\ 
\frac{\exp[-(m_1^2 t+ m_2^2 u)]}{(t+u)^{3/2}}\cdot \nonumber\\ 
 && \times \exp\left[-\left(\frac{tu}{t+u}\right){\bf k}^2\right]\
\Omega_{j}^{(1b)}({\bf k}^2;t,u)\ ,
\label{AppB.35} \end{eqnarray}
where
 
\begin{eqnarray}
\Omega_{1}^{(1b,X)}({\bf k}^2;t,u) &=& C_{NN,X}^{(I)}
\left(\frac{f_{NN\pi}}{m_\pi}\right)^2 \left(\frac{f_{NN\eta}}{m_\pi}\right)^2
\left(\frac{1}{M}\right)\cdot\left\{-\frac{15}{2}\ +
\right.\nonumber\\ &&\nonumber\\ && \left.
 -\frac{5}{2}\left(\frac{t^2-2tu+u^2}{t+u}\right){\bf k}^2 +
 tu \left(\frac{t^2 + u^2}{(t+u)^2}\right){\bf k}^4\right\}\cdot
\frac{1}{(t+u)^2}\ ,
\nonumber\\ &&\nonumber\\ 
\Omega_{4}^{(1b,X)}({\bf k}^2;t,u) &=& C_{NN,X}^{(I)}
\left(\frac{f_{NN\pi}}{m_\pi}\right)^2 \left(\frac{f_{NN\eta}}{m_\pi}\right)^2
\left(\frac{1}{M}\right)
\left\{-1+2\left(\frac{tu}{t+u}\right){\bf k}^2\right\}\cdot
\frac{1}{t+u}\ .
\label{AppB.36} \end{eqnarray}

\subsection{ Off-shell corrections in TMO-diagrams}           
\label{app:B10}
Also in this case the denominators are 
$D^{(1)}(\omega_1,\omega_2)=1/\omega_1^2\omega_2^2$ \cite{RS96a}, and
one finds that the contributions to the $\tilde{V}^{(1)}_j$ 
can be written as            
\begin{eqnarray}
\tilde{V}_{j}^{(1c)}({\bf k}) &=& 
(4\pi)^{-3/2}\ e^{m_1^2/\Lambda_1^2} e^{m_2^2/\Lambda_2^2}  
\int_{t_0}^\infty dt\ \int_{u_0}^\infty du\ 
\frac{\exp[-(m_1^2 t+ m_2^2 u)]}{(t+u)^{3/2}}\cdot \nonumber\\ 
 && \times \exp\left[-\left(\frac{tu}{t+u}\right){\bf k}^2\right]\
\Omega_{j}^{(1c)}({\bf k}^2;t,u)\ .
\label{AppB.37} \end{eqnarray}
The potentials are very similar to the pseudovector-vertex corrections
given above, see \cite{RS96a}, section V.C. We have
 
\begin{eqnarray}
\Omega_{1}^{(1c,//)}({\bf k}^2;t,u) &=& C_{NN,//}^{(I)}
\left(\frac{f_{NN\pi}}{m_\pi}\right)^2 \left(\frac{f_{NN\eta}}{m_\pi}\right)^2
\left(\frac{1}{2M}\right)\cdot\left\{-\frac{15}{2}\ +
\right.\nonumber\\ &&\nonumber\\ && \left.
 -\frac{5}{2}\left(\frac{t^2-2tu+u^2}{t+u}\right){\bf k}^2 +
 tu \left(\frac{t^2 + u^2 }{(t+u)^2}\right){\bf k}^4\right\}\cdot
\frac{1}{(t+u)^2}\ ,
\nonumber\\ &&\nonumber\\ 
\Omega_{4}^{(1c,//)}({\bf k}^2;t,u) &=& C_{NN,//}^{(I)}
\left(\frac{f_{NN\pi}}{m_\pi}\right)^2 \left(\frac{f_{NN\eta}}{m_\pi}\right)^2
\left(\frac{1}{2M}\right)
\left\{-5+2\left(\frac{tu}{t+u}\right){\bf k}^2\right\}\cdot
\frac{1}{t+u}\ . 
\label{AppB.38} \end{eqnarray}

Similarly to (\ref{AppB.32c}) it is convenient to introduce the expansion
\begin{equation}
\Omega_j^{(1a,1b,1c)}({\bf k}^2;t,u) = C_{NN,1a,1b,1c}^{(I)}
\left(\frac{f_{NN\pi}}{m_\pi}\right)^2 \left(\frac{f_{NN\eta}}{m_\pi}\right)^2
\left(\frac{1}{M}\right)\cdot
 \sum_{k=0}^K\ \Upsilon_{j,k}^{(0),1a,1b,1c}(t,u)\ \left({\bf k}^2\right)^k\ .
\label{AppB.39} \end{equation}
The functions $\Upsilon_{j,k}(t,u)$ are given in Table~\ref{table3}-\ref{table6}.\\

\section{ Partial Wave Analysis}         

\subsection{ General Structure }            
\label{sec:PWA}
We write the potentials listed in the previous section in the general form
\begin{subequations}
\label{allequations}
\begin{eqnarray}
\tilde{V}_{j}({\bf k}) &=& 
\int_{t_0}^\infty dt\ \int_{u_0}^\infty du\ w(t,u)\
 \exp\left[-\left(\frac{tu}{t+u}\right){\bf k}^2\right]\
\Omega_{j}^{(type)}({\bf k}^2;t,u)\ ,
\label{AppB.135a} \end{eqnarray}
where for the adiabatic, the $1/M$ pv- and off-shell corrections the weight function 
$w(t,u)=w_0(t,u)$ with   
\begin{eqnarray}
  w_0(t,u) &=& 
(4\pi)^{-3/2}\ e^{m_1^2/\Lambda_1^2} e^{m_2^2/\Lambda_2^2}  
\frac{\exp[-(m_1^2 t+ m_2^2 u)]}{(t+u)^{3/2}}\ ,               
\label{AppB.135b} \end{eqnarray}
and for the non-adiabatic corrections, cfrm. (\ref{AppB.33}), 
\begin{eqnarray}
  w_{na}(t,u) &=& \left(\frac{\Lambda_1^2+\Lambda_2^2}{\Lambda_1^2\Lambda_2^2}
 -(t+u)\right)\ w_0(t,u)\ .
\label{AppB.135c} \end{eqnarray}
\end{subequations}
{}From ${\bf k}^2= p_f^2+p_i^2-2p_fp_iz$, where $z=\cos\theta$, and the $\Omega_j$'s
one readily sees that the potential contributions can be written in the form
\begin{eqnarray}
\tilde{V}_{j}({\bf k}) &=& 
\int_{t_0}^\infty dt\ \int_{u_0}^\infty du\ w(t,u)\
\left(X_j(t,u) + z Y_j(t,u)+ z^2 Z_j(t,u) \vphantom{\frac{A}{A}}\right)\cdot
 \exp\left[-\left(\frac{tu}{t+u}\right){\bf k}^2\right]\ ,
\label{AppB.136} \end{eqnarray}
where
\begin{subequations}
\begin{eqnarray}
 X_j(t,u) &=& \hspace{2.3mm} \Gamma_j \left[\Upsilon_{j,0}(t,u) + \left(p_f^2+p_i^2\right)
 \Upsilon_{j,1}(t,u) + \left(p_f^2+p_i^2\right)^2\ \Upsilon_{j,2}\right]\ , 
 \label{AppB.137a}\\
 Y_j(t,u) &=& -\Gamma_j \left[ 2p_fp_i\ \Upsilon_{j,1}(t,u) 
 + 4p_fp_i \left(p_f^2+p_i^2\right)\ \Upsilon_{j,2}\right]\ , 
 \label{AppB.137b}\\
 Z_j(t,u) &=& \hspace{2.3mm}\Gamma_j \left[4p_f^2p_i^2\ \Upsilon_{j,2}\right]\ . 
 \label{AppB.137c}
 \end{eqnarray}
\end{subequations}
Here, $\Gamma_j$ is a combination of isospin and coupling constant factors.

\subsection{ Partial Wave Projection}       
\label{app:PW2}
The partial wave projection of (\ref{AppB.13}) can be done easily. Using formula 10.2.36
 of reference \cite{AS70} one derives that
\begin{eqnarray}
 e^{-\gamma {\bf k}^2} &=& 
\sum_{L=0}^\infty (2L+1)\ f_{L}(2\gamma p_f p_i)\
 P_L(z)\cdot e^{-\gamma(p_f-p_i)^2}\ \ , \ \gamma \equiv \frac{tu}{t+u}\ .
\label{AppB.15a} \end{eqnarray}
Here, the function $f_{L}(z)$ is defined as
\begin{equation}
 f_L(2\gamma p_f p_i) \equiv \sqrt{\frac{\pi}{4\gamma p_f p_i}}
 I_{L+1/2}(2\gamma p_f p_i)\cdot e^{-2\gamma p_f p_i}\ .
\label{AppB.15b} \end{equation}
We note that in the form (\ref{AppB.15a}), the gaussian damping in the 
off-shell momentum region is manifest.\\
{}From (\ref{AppB.15a}) and the recurrence relations for Legendre polynomials, one
readily obtains e.g.
\begin{eqnarray}
 z\ e^{-\gamma {\bf k}^2} &=& 
 \sum_{L=0}^\infty (2L+1) \left(\frac{L}{2L+1} f_{L-1} + 
 \frac{L+1}{2L+1} f_{L+1}\right)(2\gamma p_f p_i)\
 P_L(z)\cdot e^{-\gamma(p_f-p_i)^2}\ , \nonumber\\ &&\nonumber\\
 z^2\ e^{-\gamma {\bf k}^2} &=& 
 \sum_{L=0}^\infty (2L+1) \left(\frac{L(L-1)}{(2L+1)(2L-1)} f_{L-2} + 
 \frac{4L^3+6L^2-1}{(2L-1)(2L+1)(2L+3)} f_{L}  
 \right.\nonumber\\ && \left. +
 \frac{(L+1)(L+2)}{(2L+1)(2L+3)} f_{L+2}\right)(2\gamma p_f p_i)\
 P_L(z)\cdot e^{-\gamma(p_f-p_i)^2}\ \ . 
\label{AppB.15c} \end{eqnarray}
Using these results the partial wave contributions can be worked out in a 
straightforward manner.
The basic partial wave projections needed are
\begin{eqnarray}
     U_{L}(t,u)&=&\frac{1}{2} \int_{-1}^{+1} dz\ P_{L}(z)\ 
     \exp\left(-\gamma{\bf k}^{2}\right)\ , 
  \  R_{L}(t,u) = \frac{1}{2} \int_{-1}^{+1} dz\ z\ P_{L}(z) 
     \exp\left(-\gamma{\bf k}^{2}\right)\ , \nonumber \\
     S_{L}(t,u) &=&
     \frac{1}{2} \int_{-1}^{+1} dz\ z^2\ P_{L}(z)\ 
 \exp\left(-\gamma{\bf k}^{2}\right)\ .                
\label{projects} \end{eqnarray}
 where $\gamma = tu/(t+u)$, and the functions $U_L, R_L, S_L$ can be read off
 from (\ref{AppB.15a})-(\ref{AppB.15c}). Writing
\begin{equation}
    V({\bf p}_f,{\bf p}_i)= \sum_{j=1}^{4} \tilde{V}_{j}({\bf p}_f,{\bf p}_i)\,
                      ({\bf p}_f | P_{j} |{\bf p}_i)\ ,
\end{equation}
the partial wave expansion of the $V_{j}$-functions reads
\begin{equation}
    V_{j}({\bf p}_f,{\bf p}_i)= \sum_{L=0}^{\infty}\
(2L+1)\ \tilde{V}^{(j)}_{L}(x)\ P_{L}(\cos \theta)\ . \end{equation}
Using (\ref{AppB.15a})-(\ref{AppB.15c}) the  partial waves
$V^{(j)}_{L}(x)$ for the TME-potentials can be written as
\begin{eqnarray}
\tilde{V}_L^{(j)}(p_f,p_i) &=& 
\int_{t_0}^\infty dt\ \int_{u_0}^\infty du\ w(t,u)\
\left[X_j(t,u)\ U_L(t,u) 
 + Y_j(t,u)\ R_L(t,u) + 
 Z_j(t,u)\ S_L(t,u) \vphantom{\frac{A}{A}} \right]\ \nonumber\\
 &\equiv& {\cal S}_L\left[X_j\cdot U_L + Y_j\cdot R_L + Z_j\cdot S_L \right]\ .                   
\label{pw101} \end{eqnarray}
 
\subsection{ Partial Wave Projection Spinor Invariants}

Distinguishing between the partial waves with parity $P=(-)^{J}$
and $P=-(-)^{J}$, we write the potential matrix elements on the
LSJ-basis in the following way (see {\it e.g.}\ \cite{SNRV71},
section VII): \\
(i) $P=(-)^{J}$:
\begin{equation}
  (p_f ; L' S' J' M'|\ V\ |p_i; L S J M) =  4\pi\ V^{J,+}(S',S)
\delta_{J^{\prime}J}\ \delta_{M^{\prime}M}\ \delta_{L^{\prime}L}\ .
\label{eq:9.1}  \end{equation}
(ii) $P=-(-)^{J}$:
\begin{equation}
  (p_f ; L' S' J' M'|\ V\ |p_i; L S J M) =  4\pi\,
\delta_{J^{\prime}J}\,\delta_{M^{\prime}M}\,\delta_{S^{\prime}S}\,
 V^{J,-}(L',L)\ . 
\label{eq:9.2}  \end{equation}
For notational convenience we will use as an index the parity factor
$\eta$, which is defined by writing $ P=\eta (-)^{J}$.
The $P=(-)^{J}$-states contain the spin singlet and triplet-uncoupled
states($\eta=+$), and the $P=-(-)^{J}$-states contain
the spin triplet-coupled states ($\eta=-$).
 
In nucleon-nucleon the proton neutron mass difference $M_p-M_n$ is small,          
and, except for very special studies, the spin singlet-triplet transitions 
can be neglected. 
Actually, in the TME-potentials we take $M_p=M_n=M_N$. As a consequence
there are no anti-symmetric spin-orbit potentials, so $V_6=0$.
Also, since we restrict ourselves to terms of order up to and
including $1/M_N$, there are no contrubutions to the quadratic-spin-orbit
potentials, i.e. $V_5 = 0$. 
Also, in nucleon-nucleon one can neglect terms with $P_7$ and $P_8$ \cite{SNRV71}.
Therefore, we can restrict the partial wave
projection of the spinor invariants to the cases $P_1, P_2, P_3$, and $P_4$.

\noindent Below we list the partial wave matrix elements for $ \eta = \pm$ for
the different $V_{j}\ P_{j}, (j =1,2,3)$. Here we
restrict ourselves to the matrix elements $\neq 0$.  \\
1. {\it central} $P_{1}=1$:
\begin{eqnarray}
     (p_f ; L' S' J' M' | V^{(1)}P_{1} |p_i; L S J M) &=&
4\pi\,  \delta_{J^{\prime}J}\,\delta_{M^{\prime}M}\,
F^{J,\eta}_{1}(L^{\prime}\ S',L\ S)\ ,
    \label{PWP1} \\[0.3cm]
{\rm with } \hspace*{0.5cm}  F^{J,\eta}_{1}(L^{\prime}\ S',L\ S)\ &=&
  \delta_{L^{\prime}L}\, \delta_{S^{\prime}S}\, V^{(1)}_{L}(x) \nonumber
\end{eqnarray}
2. {\it spin-spin} $P_{2}=
\mbox{\boldmath $\sigma$}_1\cdot\mbox{\boldmath $\sigma$}_2$:
\begin{eqnarray}
     (p_f ; L' S' J' M' | V^{(2)}P_{2} |p_i; L S J M) &=&
4\pi\,  \delta_{J^{\prime}J}\,\delta_{M^{\prime}M}\,
F^{J,\eta}_{2}(L^{\prime}\ S',L\ S)\ ,
\label{PWP2} \\[0.3cm]
{\rm with} \hspace*{0.5cm}  F^{J,\eta}_{2}(L^{\prime}\ S',L\ S)\ &=&
   \delta_{L^{\prime}L}\, \delta_{S^{\prime}S}\,
   \left[2S(S+1)-3\right] V^{(2)}_{L}(x)\
\nonumber \end{eqnarray}
3. {\it tensor} $ P_{3}= (\mbox{\boldmath $\sigma$}_1\cdot{\bf k})
(\mbox{\boldmath $\sigma$}_2\cdot{\bf k}) - \frac{1}{3} 
(\mbox{\boldmath $\sigma$}_1\cdot\mbox{\boldmath $\sigma$}_2) {\bf k}^{2}$:
\begin{equation}
     (p_f ; L' S' J' M' | V^{(3)}P_{3} |p_i; L S J M) =
\frac{8\pi}{3} (p_f^{2}+p_i^{2})\,
\delta_{J^{\prime}J}\,\delta_{M^{\prime}M}\, F^{J,\eta}_{3}(i,j)\ ,
\label{eq:9.3}  \end{equation}
where $i=S'$ and $j=S$ for $\eta=+$, respectively $i=L'$ and $j=L$
for $\eta=-$. \\
(i) triplet uncoupled: $L=L'=J,\ S=S'=1$
\begin{eqnarray}
 F^{J,+}_{3}(1,1)&=&
 \left[ V^{(3)}_{J}- \frac{1}{2}\sin 2\psi
 \left(\frac{2J+3}{2J+1}V^{(3)}_{J-1}+
 \frac{2J-1}{2J+1}V^{(3)}_{J+1}\right)\right] \end{eqnarray}
(ii) triplet coupled: $L=J\pm 1,\ L'=J\pm 1,\ S=S'=1$
 \begin{subequations}
 \begin{eqnarray}
  F^{J,-}_{3}(J-1,J-1) &=& \frac{J-1}{2J+1}\ \left[ -V^{(3)}_{J-1}
  + \frac{1}{2} \sin 2\psi 
  \left\{ \frac{2J-3}{2J-1} V^{(3)}_{J}
  + \frac{2J+1}{2J-1} V^{(3)}_{J-2} \right\} \right]
 \label{eq:9.4a}\\[0.2cm]
  F^{J,-}_{3}(J-1,J+1) &=& -3 \frac{\sqrt{J(J+1)}}{2J+1}\ \left[
  - \sin 2\psi\ V^{(3)}_{J} 
 +\left(\cos^{2}\psi V^{(3)}_{J-1}
 +\sin^{2}\psi V^{(3)}_{J+1}\right) \right]
 \label{eq:9.4b}\\[0.2cm]
  F^{J,-}_{3}(J+1,J-1) &=& -3 \frac{\sqrt{J(J+1)}}{2J+1}\ \left[
  - \sin 2\psi\ V^{(3)}_{J}  
 +\left(\sin^{2}\psi V^{(3)}_{J-1}
 +\cos^{2}\psi V^{(3)}_{J+1}\right) \right]
 \label{eq:9.4c}\\[0.2cm]
  F^{J,-}_{3}(J+1,J+1) &=& \frac{J+2}{2J+1}\ \left[ -V^{(3)}_{J+1}
  + \frac{1}{2} \sin 2\psi 
  \left\{ \frac{2J+5}{2J+3} V^{(3)}_{J}
  + \frac{2J+1}{2J+3} V^{(3)}_{J+2} \right\} \right]
 \label{eq:9.4d}
\end{eqnarray}
 \end{subequations}
where we introduced
\begin{equation}
\cos\psi = \frac{p_i}{\sqrt{p_f^{2}+ p_i^{2}}} \hspace{0.5cm} , \hspace{0.5cm}
\sin\psi = \frac{p_f}{\sqrt{p_f^{2}+ p_i^{2}}}
\label{eq:9.4}  \end{equation}
4. {\it spin-orbit} $ P_{4}=\frac{i}{2}
(\mbox{\boldmath $\sigma$}_1+\mbox{\boldmath $\sigma$}_2)\cdot{\bf n}$:
\begin{eqnarray}
     (p_f ; L' S' J' M'| V^{(4)}P_{4} |p_i; L S J M)&=&
      4\pi\, p_f p_i \delta_{J^{\prime}J}\,\delta_{M^{\prime}M}\,
  F^{J,\eta}_{4}(i,j)\ .
\label{eq:9.5}  \end{eqnarray}
(i) triplet uncoupled: $L=L'=J,\ S=S'=1$
\begin{equation}
  F^{J,+}_{4}(1,1)= - \left(
         V^{(4)}_{J-1}-V^{(4)}_{J+1} \right)/(2J+1)
\label{eq:9.6}  \end{equation}
(ii) triplet coupled: $L=J\pm 1,\ L'=J\pm 1,\ S=S'=1$
\begin{eqnarray}
   F^{J,-}_{4}(J-1,J-1)&=& \hspace*{0.3cm}
    \frac{(J-1)}{(2J-1)}
    \left(V^{(4)}_{J-2}-V^{(4)}_{J}\right) \nonumber \\[0.2cm]
   F^{J,-}_{4}(J+1,J+1)&=& -
    \frac{(J+2)}{(2J+3)}\left(V^{(4)}_{J}-V^{(4)}_{J+2}\right)
\label{eq:9.7}  \end{eqnarray}

With the matrix elements of this section, the partial waves for
the potentials can be readily derived. 
Henceforth, we will use the following
shorthand notation \cite{Rijken85} for the potentials: \\
(i) $P=(-)^{J}$:
\begin{eqnarray}
     V^{J}_{0,0}  =  V^{J,+} (0,0)
     & \hspace*{0.5cm} , \hspace*{0.5cm} &
     V^{J}_{0,2}  =  V^{J,+} (0,1) \nonumber \\[0.2cm]
     V^{J}_{2,0}  =  V^{J,+} (1,0)
     & \hspace*{0.5cm} , \hspace*{0.5cm} &
     V^{J}_{2,2}  =  V^{J,+} (1,1)
\label{eq:9.8}  \end{eqnarray}
(ii) $P=-(-)^{J}$:
\begin{eqnarray}
     V^{J}_{1,1}  =  V^{J,-} (J-1,J-1)
     & \hspace*{0.5cm} , \hspace*{0.5cm} &
     V^{J}_{1,3}  =  V^{J,-} (J-1,J+1) \nonumber \\[0.2cm]
     V^{J}_{3,1}  =  V^{J,-} (J+1,J-1)
     & \hspace*{0.5cm} , \hspace*{0.5cm} &
     V^{J}_{3,3}  =  V^{J,-} (J+1,J+1)
\label{eq:9.9}  \end{eqnarray}
where it is always understood that the final and initial state
momenta are respectively $p_{f}$ and $p_{i}$.
So $V^{J}_{0,0}= V^{J}_{0,0}(p_{f},p_{i})$ etc.
Since
\begin{equation}
     V^{J}_{2,0}(p_{f},p_{i}) = V^{J}_{0,2}(p_{i},p_{f})
   \hspace*{0.5cm} , \hspace*{0.5cm}
     V^{J}_{3,1}(p_{f},p_{i}) = V^{J}_{1,3}(p_{i},p_{f})
\label{eq:9.10}  \end{equation}
we will give in case of the off-diagonal terms only the
explicit expressions for
  $V^{J}_{0,2}(p_{f},p_{i})$ and $V^{J}_{1,3}(p_{f},p_{i})$.

\subsection{ Partial Wave Projection Potentials}            
The momentum space partial wave central ($C$), spin-spin ($\sigma$),  tensor ($T$), 
and  spin-orbit ($SO$) potentials $V_L^{(C)}=V^{(1)}_L(p_f,p_i)$ etc. 
lead to the following partial wave potentials $\neq 0$:

 \begin{subequations}
\begin{eqnarray}
 V_{0,0}^J &=& 4\pi \left( V_J^{(C)} - 3 V_J^{(\sigma)}\right) 
 \label{eq:9.11a}\\
 V_{2,2}^J &=& 4\pi \left[\left( V_J^{(C)}+V_J^{(\sigma)}\right) 
 +\frac{2}{3}\left(p_f^2+p_i^2\right) \left\{ V_J^{(T)} -
  \frac{1}{2}\sin 2\psi \left(\frac{2J+3}{2J+1}\ V_{J-1}^{(T)}
 +\frac{2J-1}{2J+1}\ V_{J+1}^{(T)}\right)\right\} \right.\nonumber\\
 && \hspace{5mm} \left. - p_f p_i \left(V^{(SO)}_{J-1}-V^{(SO)}_{J+1}\right)/(2J+1)
 \vphantom{\frac{A}{A}}\right] \label{eq:9.11b}\\
 V_{1,1}^J &=& 4\pi \left[\left( V_{J-1}^{(C)}+V_{J-1}^{(\sigma)}\right) 
 +\frac{2}{3}\left(p_f^2+p_i^2\right) \frac{J-1}{2J+1}\left\{ -V_{J-1}^{(T)} 
 +\frac{1}{2}\sin 2\psi \left(\frac{2J-3}{2J-1}\ V_J^{(T)}
 +\frac{2J+1}{2J-1}\ V_{J-2}^{(T)}\right)\right\}\right. \nonumber\\
 && \hspace{5mm} \left. + p_f p_i(J-1)\left(V^{(SO)}_{J-2}-V^{(SO)}_J\right)/(2J-1)
 \vphantom{\frac{A}{A}}\right] \label{eq:9.11c}\\
 V_{1,3}^J &=& 4\pi \left[ 2\left(p_f^2+p_i^2\right) 
 \frac{\sqrt{J(J+1)}}{2J+1} \left\{ \vphantom{\frac{A}{A}}
 \sin 2\psi\ V_J^{(T)} 
  -\left(\cos^2\psi\ V^{(T)}_{J-1} + \sin^2\psi\ V^{(T)}_{J+1}\right)
 \right\}\right] \label{eq:9.11d}\\
 V_{3,3}^J &=& 4\pi \left[\left( V_{J+1}^{(C)}+V_{J+1}^{(\sigma)}\right) 
 +\frac{2}{3}\left(p_f^2+p_i^2\right) \frac{J+2}{2J+1}\left\{ -V_{J+1}^{(T)} 
 +\frac{1}{2}\sin 2\psi \left(\frac{2J+5}{2J+3}\ V_{J}^{(T)}
 +\frac{2J+1}{2J+3}\ V_{J+2}^{(T)}\right)\right\}\right.\nonumber\\     
 && \hspace{5mm} \left. - p_f p_i(J+2)\left(V^{(SO)}_J-V^{(SO)}_{J+2}\right)/(2J+3)
 \vphantom{\frac{A}{A}}\right]
\label{eq:9.11e}  \end{eqnarray}
 \end{subequations}
Notice that $V_{0,2}^J= V_{2,0}^J=0$ because $V_L^{(ASO)}=0$. Furthermore,
$V_{3,1}^J= V_{1,3}^J$.                                        

\section{ Adiabatic PS-PS Potentials}                
\label{sec:adPSPS}
 
\subsection{ Adiabatic PS-PS Coefficients}               
Defining the shorthands $A(p_f,p_i)$ and $B(p_f,p_i)$ by
\begin{equation}
 {\bf k}^2= (p_f^2+p_i^2)-2 p_f p_i z \equiv A-Bz\ \ , \ \
 {\bf k}^4 \equiv A^2 - 2 A B z + B^2 z^2\ .                      
\label{eq:8.1} \end{equation}
one obtains from (\ref{AppB.32a}) for the coefficients $X,Y,Z \neq 0$: \\

\noindent (i)\ central:
\begin{eqnarray}
 && X^{(//)}_1(t,u) = +\frac{2}{\sqrt{\pi}}\ C_{NN,//}^{(I)}
\left(\frac{f_{NN\pi}}{m_\pi}\right)^2 \left(\frac{f_{NN\eta}}{m_\pi}\right)^2\cdot
\nonumber\\ &&\nonumber\\ && \times
 \frac{1}{4}\left\{ 15 + 
 2\left(\frac{t^2-8ut+u^2}{t+u}\right) A(p_f,p_i)+ 4\left(\frac{t^2u^2}{(t+u)^2}\right)
 A^2(p_f,p_i)\right\}\cdot\frac{\sqrt{t+u}}{(t+u)^2}\ ,
\nonumber\\ &&\nonumber\\ 
 && Y^{(//)}_1(t,u) = -\frac{1}{\sqrt{\pi}}\ C_{NN,//}^{(I)}
\left(\frac{f_{NN\pi}}{m_\pi}\right)^2 \left(\frac{f_{NN\eta}}{m_\pi}\right)^2\cdot
\nonumber\\ &&\nonumber\\ && \times \left\{  
 \left(\frac{t^2-8ut+u^2}{t+u}\right) B(p_f,p_i)+ 4\left(\frac{t^2u^2}{(t+u)^2}\right)
 A(p_f,p_i) B(p_f,p_i)\right\}\cdot\frac{\sqrt{t+u}}{(t+u)^2}\ ,
\nonumber\\ &&\nonumber\\ 
 && Z^{(//)}_1(t,u) = +\frac{2}{\sqrt{\pi}}\ C_{NN,//}^{(I)}
\left(\frac{f_{NN\pi}}{m_\pi}\right)^2 \left(\frac{f_{NN\eta}}{m_\pi}\right)^2\cdot
\nonumber\\ &&\nonumber\\ && \times
 \left(\frac{t^2u^2}{(t+u)^2}\right) B^2(p_f,p_i)\cdot\frac{\sqrt{t+u}}{(t+u)^2}\ ,
\label{eq:8.2} \end{eqnarray}

\noindent (ii)\ spin-spin:
\begin{eqnarray}
&& X^{(//)}_2(t,u) = -\frac{2}{3\sqrt{\pi}}\ C_{NN,//}^{(I)}
\left(\frac{f_{NN\pi}}{m_\pi}\right)^2 \left(\frac{f_{NN\eta}}{m_\pi}\right)^2
\cdot A(p_f,p_i)\cdot\frac{\sqrt{t+u}}{(t+u)}\ ,
\nonumber\\ &&\nonumber\\
&& Y^{(//)}_2(t,u) = +\frac{2}{3\sqrt{\pi}}\ C_{NN,//}^{(I)}
\left(\frac{f_{NN\pi}}{m_\pi}\right)^2 \left(\frac{f_{NN\eta}}{m_\pi}\right)^2
\cdot B(p_f,p_i)\cdot\frac{\sqrt{t+u}}{(t+u)}\ ,
\label{eq:8.3} \end{eqnarray}

\noindent (iii)\ tensor:     
\begin{eqnarray}
&& X^{(//)}_3(t,u) = +\frac{1}{\sqrt{\pi}}\ C_{NN,//}^{(I)}
\left(\frac{f_{NN\pi}}{m_\pi}\right)^2 \left(\frac{f_{NN\eta}}{m_\pi}\right)^2
\cdot \frac{\sqrt{t+u}}{(t+u)}\ .
\label{eq:8.4} \end{eqnarray}
For the $X$-diagram contributions one has
\begin{eqnarray}
 X^{(X)}_1 &=& - X^{(//)}_1\ ,\ \ Y^{(X)}_1  =  - Y^{(//)}_1\ ,\ \ 
 Z^{(X)}_1  =  - Z^{(//)}_1\ , \nonumber\\
 X^{(X)}_2 &=& + X^{(//)}_2\ ,\ \ Y^{(X)}_2  =  + Y^{(//)}_2\ , \nonumber\\
 X^{(X)}_3 &=& + X^{(//)}_3\ .              
\label{eq:8.5} \end{eqnarray}
 
\subsection{ Adiabatic PS-PS Partial Wave Potentials}                
The central, spin-spin, and tensor partial wave contributions are 
\begin{eqnarray}
 V^{(C)}_L(p_f,p_i) &=& {\cal S}_L\left[ X_1\cdot U_L+Y_1\cdot R_L+Z_1\cdot S_L\right] 
\equiv {\cal S}_L\left[{\bf X}_C\cdot{\bf U}_L\right]\ , \nonumber\\
 V^{(\sigma)}_L(p_f,p_i) &=& {\cal S}_L\left[ X_2\cdot U_L+Y_2\cdot R_L+Z_2\cdot S_L\right] 
\equiv {\cal S}_L\left[{\bf X}_\sigma\cdot{\bf U}_L\right]\ , \nonumber\\
 V^{(T)}_L(p_f,p_i) &=& {\cal S}_L\left[ X_3\cdot U_L\right]\ ,
\label{eq:8.6} \end{eqnarray}
where
\begin{eqnarray}
 X_C(t,u) &=& X^{(//)}_1 + X^{(X)}_1\ \ ,\ \
 X_\sigma(t,u) = X^{(//)}_2 + X^{(X)}_2\ , \nonumber\\
 X_T(t,u) &=& X^{(//)}_3 + X^{(X)}_3\ ,
\label{eq:8.7} \end{eqnarray}
and similar formulas for $Y_{C,\sigma,T}$ and $Z_{C,\sigma,T}$.
 
The momentum space partial wave central, spin-spin, and tensor potentials
$V^{(C)}_L(p_f,p_i)$ etc. lead to the following contributions:          
\begin{subequations}
\begin{eqnarray}
 V_{0,0}^J(ad) &=& 4\pi \left( V_J^{(C)} - 3 V_J^{(\sigma)}\right) (p_f,p_i)
 \label{eq:8.8a}\\
 V_{2,2}^J(ad) &=& 4\pi \left[\left( V_J^{(C)}+V_J^{(\sigma)}\right) 
 +\frac{2}{3}\left(p_f^2+p_i^2\right) \left\{ V_J^{(T)} -
  \frac{1}{2}\sin 2\psi \left(\frac{2J+3}{2J+1}\ V_{J-1}^{(T)}
 +\frac{2J-1}{2J+1}\ V_{J+1}^{(T)}\right)\right\}\right] \label{eq:8.8b}\\
 V_{1,1}^J(ad) &=& 4\pi \left[\left( V_{J-1}^{(C)}+V_{J-1}^{(\sigma)}\right) 
 +\frac{2}{3}\left(p_f^2+p_i^2\right) \frac{J-1}{2J+1}\left\{ -V_{J-1}^{(T)} 
 +\frac{1}{2}\sin 2\psi \left(\frac{2J-3}{2J-1}\ V_{J}^{(T)}
 +\frac{2J+1}{2J-1}\ V_{J-2}^{(T)}\right)\right\}\right] \label{eq:8.8c}\\
 V_{1,3}^J(ad) &=& 4\pi \left[ 2\left(p_f^2+p_i^2\right) 
 \frac{\sqrt{J(J+1)}}{2J+1} \left\{ \vphantom{\frac{A}{A}}
 \sin 2\psi\ V_J^{(T)} 
  -\left(\cos^2\psi\ V^{(T)}_{J-1} + \sin^2\psi\ V^{(T)}_{J+1}\right)
 \right\}\right] \label{eq:8.8d}\\
 V_{3,3}^J(ad) &=& 4\pi \left[\left( V_{J+1}^{(C)}+V_{J+1}^{(\sigma)}\right) 
 +\frac{2}{3}\left(p_f^2+p_i^2\right) \frac{J+2}{2J+1}\left\{ -V_{J+1}^{(T)} 
 +\frac{1}{2}\sin 2\psi \left(\frac{2J+5}{2J+3}\ V_{J}^{(T)}
 +\frac{2J+1}{2J+3}\ V_{J+2}^{(T)}\right)\right\}\right]                
\label{eq:8.8e} \end{eqnarray}
\end{subequations}

\end{widetext}

\appendix

\section{ Miscellaneous  Integrals}        
\label{app:C}
In this appendix we list a number of useful integrals.\\

 \noindent (i)\ Consider the integrals with $p$-components of the 
 $\mbox{\boldmath $\Delta$}$-vector in the integrand:
gaussian integral
\begin{eqnarray}
 {\cal I}^{(p)}_{m, \ldots, n}({\bf k}) &=& 
 \int \frac{d^3 \Delta}{(2\pi)^3}\ \Delta_m \ldots \Delta_n\
 \exp\left[-a \mbox{\boldmath $\Delta$}^2 + 
 2b{\bf k}\cdot\mbox{\boldmath $\Delta$}\right] \nonumber\\ &=&
 (4\pi a)^{-3/2}\ (2b)^{-p}\ \tilde{\nabla}_m \ldots \tilde{\nabla}_l\
 \exp\left[+\frac{b^2}{a} {\bf k}^2\right] \nonumber\\
 &\equiv & (4\pi a)^{-3/2}\ \Pi^{(p)}_{m \ldots n}({\bf k}^2)\ 
 \exp\left[+\frac{b^2}{a} {\bf k}^2\right]\ .
\label{AppC.1} \end{eqnarray}
The first tensors $\Pi^{(p)}_{m \ldots n}$ are found to be
\begin{eqnarray}
 \Pi^{(0)} &=& 1\ ,\  \Pi^{(1)}_{m} = \frac{b}{a}\ k_m\ ,\ \Pi^{(2)}_{mn} = \frac{1}{a}
 \left\{\frac{1}{2}\delta_{mn}+\frac{b^2}{a} k_m k_n\right\}\ ,
 \nonumber\\
 \Pi^{(3)}_{mnk} &=& \frac{b}{a^2}\left\{ \frac{1}{2}\left[\delta_{mn} k_k +            
 \delta_{mk} k_n + \delta_{nk} k_m \right] +\frac{b^2}{a} k_m k_n k_k\right\}\ ,
 \nonumber\\
 \Pi^{(4)}_{mnkl} &=& \frac{1}{a^2}\left\{  \frac{1}{4}\left[
\delta_{mn}\delta_{kl} + \delta_{mk}\delta_{nl} + \delta_{nk}\delta_{ml}\right] 
 +\frac{1}{2}\frac{b^2}{a}
 \cdot\right.\nonumber\\ && \left.\times
 \left[ \delta_{mn} k_k k_l +\delta_{mk} k_n k_l       
 +\delta_{nk} k_m k_l  
 \right.\right.\nonumber\\ &&  \hspace{1cm} \left.\left. 
 +\delta_{kl} k_m k_n +\delta_{ml} k_k k_n        
 +\delta_{nl} k_k k_m \right]
 \right.\nonumber\\ &&  \hspace{1cm} \left. 
 +\left(\frac{b^2}{a}\right)^2 k_m k_n k_k k_l\right\}\ .
\label{AppC.2} \end{eqnarray}

 \noindent (ii)\ Consider the integrals with $p$-factors           
 $\mbox{\boldmath $\Delta$}^2$ in the integrand:
\begin{eqnarray}
 {\cal J}^{(p)}({\bf k}) &=& \int \frac{d^3 \Delta}{(2\pi)^3}\ 
 (\mbox{\boldmath $\Delta$}^2)^p\ 
 \exp\left[-a \mbox{\boldmath $\Delta$}^2 + 
 2b{\bf k}\cdot\mbox{\boldmath $\Delta$}\right] \nonumber\\ &=&
 \left(-\frac{\partial}{\partial a}\right)^p\cdot(4\pi a)^{-3/2}
 \exp\left[+\frac{b^2}{a} {\bf k}^2\right] \nonumber\\
 &\equiv & (4\pi a)^{-3/2}\ \Gamma^{(p)}({\bf k}^2)\ 
 \exp\left[+\frac{b^2}{a} {\bf k}^2\right]\ .
\label{AppC.3} \end{eqnarray}
The first tensors $\Gamma^{(p)}_{m \ldots n}$ are 
\begin{eqnarray}
 \Gamma^{(0)} &=& 1\ ,\  \Gamma^{(1)} = \frac{1}{2a}\left[ 3             
 +2a\left(\frac{b}{a}\right)^2 {\bf k}^2\right]\ , \nonumber\\ 
 \Gamma^{(2)} &=& \frac{1}{4a^2}\left[ 15 + 20a\left(\frac{b}{a}\right)^2
 {\bf k}^2 +4a^2\left(\frac{b}{a}\right)^4 {\bf k}^4\right]\ , \nonumber\\
 \Gamma^{(3)} &=& \frac{1}{8a^3}\left[ 105 + 210 a\left(\frac{b}{a}\right)^2
 {\bf k}^2 +84a^2\left(\frac{b}{a}\right)^4 {\bf k}^4
 \right.\nonumber\\ && \hspace{1cm} \left.
  +8 a^3\left(\frac{b}{a}\right)^6 {\bf k}^6 \right]\ .                  
\label{AppC.4} \end{eqnarray}

 \noindent (iii)\ Consider the integrals with $p$-factors           
 $\mbox{\boldmath $\Delta$}\cdot{\bf k}$ in the integrand:
\begin{eqnarray}
 {\cal K}^{(p)}({\bf k}) &=& \int \frac{d^3 \Delta}{(2\pi)^3}\ 
 (\mbox{\boldmath $\Delta$}\cdot{\bf k})^p\ 
 \exp\left[-a \mbox{\boldmath $\Delta$}^2 + 
 2b{\bf k}\cdot\mbox{\boldmath $\Delta$}\right] \nonumber\\ &=&
 (4\pi a)^{-3/2}\ \left(\frac{1}{2}\frac{\partial}{\partial b}\right)^p\
 \exp\left[+\frac{b^2}{a} {\bf k}^2\right] \nonumber\\
 &\equiv & (4\pi a)^{-3/2}\ \Sigma^{(p)}({\bf k}^2)\ 
 \exp\left[+\frac{b^2}{a} {\bf k}^2\right]\ .
\label{AppC.5} \end{eqnarray}
The first coefficients $\Sigma^{(p)}$ are                  
\begin{eqnarray}
 \Sigma^{(0)} &=& 1\ ,\  \Sigma^{(1)} = \frac{b}{a}\ {\bf k}^2\ , \
 \nonumber\\ 
 \Sigma^{(2)} &=& \frac{1}{2a}
 \left[1+2a\left(\frac{b}{a}\right)^2 {\bf k}^2\right]{\bf k}^2\ ,             
 \nonumber\\ 
 \Sigma^{(3)} &=& \frac{b}{2a^2}
 \left[3+2a\left(\frac{b}{a}\right)^2 {\bf k}^2\right]{\bf k}^4\ .             
\label{AppC.6} \end{eqnarray}

 \noindent (iv)\ Consider finally the more general integrals with $p$-factors           
 $\mbox{\boldmath $\Delta$}^2$ and $q$-factors  
$\mbox{\boldmath $\Delta$}\cdot{\bf k}$ in the integrand:
\begin{eqnarray}
 {\cal J}^{(p,q)}({\bf k}) &=& \int\frac{d^3 \Delta}{(2\pi)^3} 
 (\mbox{\boldmath $\Delta$}^2)^p 
 (\mbox{\boldmath $\Delta$}\cdot{\bf k})^q                      
 \exp\left[-a \mbox{\boldmath $\Delta$}^2 + 
 2b{\bf k}\cdot\mbox{\boldmath $\Delta$}\right] \nonumber\\ &=&
 \left(-\frac{\partial}{\partial a}\right)^p
 \left(\frac{1}{2}\frac{\partial}{\partial b}\right)^q
 \cdot(4\pi a)^{-3/2}
 \exp\left[+\frac{b^2}{a} {\bf k}^2\right] \nonumber\\
 &\equiv & (4\pi a)^{-3/2}\ \Gamma^{(p,q)}({\bf k}^2)\ 
 \exp\left[+\frac{b^2}{a} {\bf k}^2\right]\ . \nonumber\\
\label{AppC.7} \end{eqnarray}
The first tensors $\Gamma^{(p,q)}_{m \ldots n}$, with $p,q \neq 0$, are 
\begin{eqnarray}
 \Gamma^{(1,1)} &=& \frac{1}{2a}\left(\frac{b}{a}\right)\left[ 5             
 +2a\left(\frac{b}{a}\right)^2 {\bf k}^2\right]{\bf k}^2\ , \nonumber\\ 
 \Gamma^{(1,2)} &=& \frac{1}{4a^2}\left[ 5 + 16a\left(\frac{b}{a}\right)^2
 {\bf k}^2 +4a^2\left(\frac{b}{a}\right)^4 {\bf k}^4\right]{\bf k}^2\ , \nonumber\\
 \Gamma^{(2,1)} &=& \frac{1}{4a^2}\left(\frac{b}{a}\right) 
 \left[ 35 + 28 a\left(\frac{b}{a}\right)^2 {\bf k}^2    
 +4a^2\left(\frac{b}{a}\right)^4 {\bf k}^4\right] {\bf k}^2\ . \nonumber\\
\label{AppC.8} \end{eqnarray}

\begin{widetext}

\section{ Integration Dictionary}         
\label{app:D}
In this appendix we give a dictionary for the evaluation of the momentum
integrals that occur in the matrix elements of the TME-potentials.
The results of the $d^3\Delta$-integration are given apart from a factor
$(4\pi a)^{-3/2}\ (a= t+u)$, common to all integrals. Using the results given in
Appendix~\ref{app:C} one obtains:

\begin{eqnarray}
 && a.\ ({\bf k}_1\cdot{\bf k}_2)^2 = \left(\mbox{\boldmath $\Delta$}\cdot{\bf k}
 -\mbox{\boldmath $\Delta$}^2\right)^2 \nonumber\\ 
 &&\Rightarrow \frac{1}{4}\left\{ 15 + 
 2\left(\frac{t^2-8ut+u^2}{t+u}\right){\bf k}^2 + 4\left(\frac{t^2u^2}{(t+u)^2}\right)
 {\bf k}^4\right\}\cdot\frac{1}{(t+u)^2}\ ,
\label{AppD.1} \\ && \nonumber\\  
 && b.\ \left[\mbox{\boldmath $\sigma$}_1\cdot{\bf k}_1\times{\bf k}_2\right] 
 \left[\mbox{\boldmath $\sigma$}_2\cdot{\bf k}_1\times{\bf k}_2\right] =
 \left[\mbox{\boldmath $\sigma$}_1\cdot\mbox{\boldmath $\Delta$}\times{\bf k}\right]
 \left[\mbox{\boldmath $\sigma$}_2\cdot\mbox{\boldmath $\Delta$}\times{\bf k}\right]
 \nonumber\\ && \Rightarrow \frac{1}{2}\left\{\frac{2}{3}
 \left(\mbox{\boldmath $\sigma$}_1\cdot\mbox{\boldmath $\sigma$}_2\right)\
{\bf k}^2\ - \left[
 \left(\mbox{\boldmath $\sigma$}_1\cdot{\bf k}\right)\
 \left(\mbox{\boldmath $\sigma$}_2\cdot{\bf k}\right) - \frac{1}{3}
 \left(\mbox{\boldmath $\sigma$}_1\cdot\mbox{\boldmath $\sigma$}_2\right)\
 {\bf k}^2\right]\right\}\cdot \frac{1}{t+u}\ ,              
\label{AppD.2} \\  && \nonumber\\   
 && c.\ [\left(\mbox{\boldmath $\sigma$}_1+\mbox{\boldmath $\sigma$}_2\right)\cdot
 {\bf k}_1\times{\bf k}_2]\ {\bf q}\cdot\left({\bf k}_1-{\bf k}_2\right) =
 [\left(\mbox{\boldmath $\sigma$}_1+\mbox{\boldmath $\sigma$}_2\right)\cdot
 \mbox{\boldmath $\Delta$}\times{\bf k}]\ 
 {\bf q}\cdot\left(\mbox{\boldmath $2\Delta$}-{\bf k}\right) 
\nonumber\\ && \Rightarrow 
 [\left(\mbox{\boldmath $\sigma$}_1+\mbox{\boldmath $\sigma$}_2\right)\cdot
 {\bf q}\times{\bf k}]\cdot\frac{1}{t+u}\ ,
\label{AppD.3} \\  && \nonumber\\ 
 && d.\ \left({\bf k}_1\cdot{\bf k}_2\right)\left({\bf k}_1^2+{\bf k}_2^2\right) =
 -2\left( \mbox{\boldmath $\Delta$}\cdot{\bf k}-
 \mbox{\boldmath $\Delta$}^2\right)^2 + \left(
 \mbox{\boldmath $\Delta$}\cdot{\bf k}-\mbox{\boldmath $\Delta$}^2\right)  
 {\bf k}^2 \nonumber\\ && \Rightarrow -\frac{1}{2} 
 \left\{15 +5\left(\frac{t^2-2tu+u^2}{t+u}\right){\bf k}^2 -
 2tu\left(\frac{t^2 + u^2}{(t+u)^2}\right){\bf k}^4\right\}\cdot
\frac{1}{(t+u)^2}\ ,
\label{AppD.4} \\ && \nonumber\\   
 && e.\ \left({\bf k}_1\cdot{\bf k}_2\right)
 = \mbox{\boldmath $\Delta$}\cdot{\bf k} -\mbox{\boldmath $\Delta$}^2 \Rightarrow
\frac{1}{2}\left\{-3+2\left(\frac{tu}{t+u}\right){\bf k}^2\right\}\cdot
\frac{1}{t+u}\ .
\label{AppD.5} \\ && \nonumber\\  
 && f.\ ({\bf k}_1\cdot{\bf k}_2)^3 = \left(\mbox{\boldmath $\Delta$}\cdot{\bf k}
 -\mbox{\boldmath $\Delta$}^2\right)^3 \nonumber\\ 
 &&\Rightarrow -\frac{1}{8}
 \left\{ 105 + 30\left(\frac{t^2-5ut+u^2}{t+u}\right){\bf k}^2 
 -12tu \left(\frac{t^2-5ut+u^2}{(t+u)^2}\right){\bf k}^4 
 -8\left(\frac{t^3u^3}{(t+u)^3}\right) {\bf k}^6\right\}\cdot\frac{1}{(t+u)^3}\ ,
\label{AppD.6} \\ && \nonumber\\  
 && g.\ ({\bf k}_1\cdot{\bf k}_2)
\left[\mbox{\boldmath $\sigma$}_1\cdot{\bf k}_1\times{\bf k}_2\right] 
 \left[\mbox{\boldmath $\sigma$}_2\cdot{\bf k}_1\times{\bf k}_2\right] =
 (\mbox{\boldmath $\Delta$}\cdot{\bf k} -\mbox{\boldmath $\Delta$}^2)
 \left[\mbox{\boldmath $\sigma$}_1\cdot\mbox{\boldmath $\Delta$}\times{\bf k}\right]
 \left[\mbox{\boldmath $\sigma$}_2\cdot\mbox{\boldmath $\Delta$}\times{\bf k}\right]
 \nonumber\\ && \Rightarrow -\frac{1}{2}\left\{\vphantom{\frac{A}{A}}
 \left(\mbox{\boldmath $\sigma$}_1\cdot\mbox{\boldmath $\sigma$}_2\right){\bf k}^2\ - 
 \left(\mbox{\boldmath $\sigma$}_1\cdot{\bf k}\right)\
 \left(\mbox{\boldmath $\sigma$}_2\cdot{\bf k}\right)\right\}
 \cdot \left[\frac{5}{2}-\frac{tu}{t+u}{\bf k}^2\right]\cdot\frac{1}{(t+u)^2}\ ,              
\label{AppD.7}
\end{eqnarray}
\begin{eqnarray}
 && h.\ ({\bf k}_1\cdot{\bf k}_2)
[\left(\mbox{\boldmath $\sigma$}_1+\mbox{\boldmath $\sigma$}_2\right)\cdot
 {\bf k}_1\times{\bf k}_2]\ {\bf q}\cdot\left({\bf k}_1-{\bf k}_2\right) =
 (\mbox{\boldmath $\Delta$}\cdot{\bf k} -\mbox{\boldmath $\Delta$}^2)
 [\left(\mbox{\boldmath $\sigma$}_1+\mbox{\boldmath $\sigma$}_2\right)\cdot
 \mbox{\boldmath $\Delta$}\times{\bf k}]\ 
 {\bf q}\cdot\left(\mbox{\boldmath $2\Delta$}-{\bf k}\right) 
\nonumber\\ && \Rightarrow -\frac{1}{2}
 [\left(\mbox{\boldmath $\sigma$}_1+\mbox{\boldmath $\sigma$}_2\right)\cdot
 {\bf q}\times{\bf k}]\cdot
 \left[5-2\frac{tu}{t+u}{\bf k}^2\right]\cdot\frac{1}{(t+u)^2}\ .              
\label{AppD.8} \end{eqnarray}

\section{On the LSJ-representation Operators}
The spherical wave functions in momentum space with quantum numbers
J, L, S, are in the SYM-convention \cite{SYM57}
\begin{equation}
  {\cal Y}_{J L S}^{M}(\hat{\bf p})= i^{L}\
  \mbox{\large\bf $C$}^{J\ L\ S}_{M\ m\ \mu}
  Y^{L}_{m}(\hat{\bf p}) \chi^{S}_{\mu} 
\label{eq:appb.1} \end{equation}
where $\chi$ is the two-nucleon spin wave function \cite{SYM}. Then
\begin{eqnarray}
    \left({\bf S}\cdot\hat{\bf p}\right) {\cal Y}_{J L S}^{M}(\hat{\bf p})&=&
 -\sqrt{6}\ i\ (-)^{L} \left\{ \sqrt{\frac{L}{2L-1}}
 \left[\begin{array}{ccc}
       L & S & J \\
       1 & 1 & 0 \\
     L-1 & S & J
        \end{array}\right] {\cal Y}_{J L-1 S}^{M}(\hat{\bf p}) \right.
  \nonumber \\ & & \nonumber \\ & & \left. \hspace*{1.8cm}
  + \sqrt{\frac{L+1}{2L+3}}
 \left[\begin{array}{ccc}
       L & S & J \\
       1 & 1 & 0 \\
     L+1 & S & J
        \end{array}\right] {\cal Y}_{J L+1 S}^{M}(\hat{\bf p}) \right\}
\nonumber \\ 
\label{eq:appb.2} \end{eqnarray}
where the $9j$-symbols differ from \cite{Edmonds57}, formula
$(6.4.4)$, in the replacement of the $3j$-symbols by the
Clebsch-Gordan coefficients and by leaving out the $m_{33}$-summation
(see \cite{Somers}). Working this out explicitly, we find
\begin{eqnarray}
    \left({\bf S}\cdot\hat{\bf p}\right) {\cal Y}_{J J-1 1}^{M}(\hat{\bf p})&=&
    -i\ a_{J}\ {\cal Y}_{J J 1}^{M}(\hat{\bf p}) \nonumber \\
    \left({\bf S}\cdot\hat{\bf p}\right) {\cal Y}_{J J+1 1}^{M}(\hat{\bf p})&=&
  \hspace*{0.3cm} i\ b_{J}\ {\cal Y}_{J J 1}^{M}(\hat{\bf p}) \\
    \left({\bf S}\cdot\hat{\bf p}\right) {\cal Y}_{J\ J\ 1}^{M}(\hat{\bf p})&=&
  \hspace*{0.3cm}  i\ a_{J}\ {\cal Y}_{J J-1 1}^{M}(\hat{\bf p}) -i\
    b_{J}\ {\cal Y}_{J J+1 1}^{M}(\hat{\bf p})\ ,            
\label{eq:appb.3} \end{eqnarray}
where
\begin{equation}
  a_{J}= -\sqrt{\frac{J+1}{2J+1}} \hspace*{1.0cm} ,
  \hspace*{1.0cm} b_{J}= -\sqrt{\frac{J}{2J+1}} 
\label{eq:appb.4} \end{equation}
 Ordering the states according to $L= J-1, L=J, L=J+1$ ,
we can write in matrix form \\
\begin{eqnarray}
  \left(\begin{array}{ccc}
   L&=&J-1 \\  & &J \\  & & J+1 \end{array}\right\|
  {\bf S}\cdot\hat{\bf p}
  \left\|\begin{array}{ccc}
  L&=& J-1 \\ & & J \\ & & J+1 \end{array}\right) =
  \left(\begin{array}{ccc}
  0 & i a_{J} & 0 \\ -i a_{J} & 0 & i b_{J} \\ 0 & -i b_{J} & 0
  \end{array} \right) 
\label{eq:appb.5} \end{eqnarray}
Similarly, using for $-i (\hat{\bf p}_f\times\hat{\bf p}_i)\cdot{\bf S}$ for
sperical components the formula
\begin{equation}
 -i (\hat{\bf p}_f\times\hat{\bf p}_i)_{n}=-\frac{4\pi}{3}\sqrt{2}\
 \mbox{\large\bf $C$}^{1 1 1}_{k l n}
 Y^{1}_{k}(\hat{\bf p}_f)Y^{1}_{l}(\hat{\bf p}_i)
\label{eq:appb.6} \end{equation}
one can work out the partial wave matrix elements involving this
operator. \\
{}From the results above one can derive the following useful
partial wave projections for the spin triplet states:
\begin{subequations}
\begin{eqnarray}
 ( L' 1 J| V({\bf k}^{2})\left({\bf S}\cdot\hat{\bf p}_i\right)^{2}
      | L 1 J )&=& 4\pi
  \left(\begin{array}{ccc}
  a_{J}^{2}V_{J-1} & 0 & - a_{J}b_{J}V_{J-1} \\[0.2cm]
   0 & V_{J}& 0 \\[0.2cm]
  - a_{J}b_{J}V_{J+1} & 0 & b_{J}^{2}V_{J+1}   \end{array}
  \right) \label{eq:appb.7a} \\[0.3cm]
 ( L' 1 J|({\bf S}\cdot\hat{\bf p}_f)^{2} V({\bf k}^{2})
      | L 1 J )&=& 4\pi
  \left(\begin{array}{ccc}
  a_{J}^{2}V_{J-1} & 0 & - a_{J}b_{J}V_{J+1} \\[0.2cm]
   0 & V_{J}& 0 \\[0.2cm]
  - a_{J}b_{J}V_{J-1} & 0 & b_{J}^{2}V_{J+1}   \end{array}
  \right) \label{eq:appb.7b} \\[0.3cm]
 ( L' 1 J|({\bf S}\cdot\hat{\bf p}_f) V({\bf k}^{2})
 ({\bf S}\cdot\hat{\bf p}_i) | L 1 J ) &=&
4\pi
  \left(\begin{array}{ccc}
  a_{J}^{2}V_{J} & 0 & - a_{J}b_{J}V_{J} \\[0.2cm]
  0 &  a_{J}^{2}V_{J-1}+b_{J}^{2}V_{J+1}& 0 \\[0.2cm]
  - a_{J}b_{J}V_{J} & 0 & b_{J}^{2}V_{J}   \end{array}
  \right)\ ,           
\label{eq:appb.7c} \end{eqnarray}
\end{subequations}
 and
\begin{equation}
  ( L' 1 J|-i(\hat{\bf p}_f\times\hat{\bf p}_i)\cdot{\bf S}\,
  V({\bf k}^{2}) | L 1 J ) =  \frac{4\pi}{2J+1} \left\{
         \begin{array}{ccl}
    (J-1)\left(V_{J-2}-V_{J}\right) & , & L=L'= J-1 \\[0.2cm]
   -\left(V_{J-1}-V_{J+1}\right) & , & L=L'= J \\[0.2cm]
    -(J+2)\left(V_{J}-V_{J+2}\right) & , & L=L'= J+1 \end{array}
  \right.                 
\label{eq:appb.8} \end{equation}
Using the identity
\begin{equation}
(\mbox{\boldmath $\sigma$}_1\cdot{\bf a})(\mbox{\boldmath $\sigma$}_2\cdot{\bf a})=
      2({\bf S}\cdot{\bf a})^{2} - {\bf a}^{2} 
\label{eq:appb.9} \end{equation}
the tensor operator can be written as
\begin{eqnarray}
 P_{3} &=&
(\mbox{\boldmath $\sigma$}_1\cdot{\bf k})(\mbox{\boldmath $\sigma$}_2\cdot{\bf k})
-\frac{1}{3}(\mbox{\boldmath $\sigma$}_1\cdot\mbox{\boldmath $\sigma$}_2){\bf k}^2 =
 \frac{1}{3} \left[ p_i^{2}\, S_{12}(\hat{\bf p}_i) +
 p_f^{2}\, S_{12}(\hat{\bf p}_f) \right] \nonumber \\[0.2cm]  & &
   -  4\left({\bf S}\cdot{\bf p}_f\right)\left({\bf S}\cdot{\bf p}_i\right)
   +  2 i \left({\bf p}_f\times{\bf p}_i\right)\cdot{\bf S}
   +  \frac{4}{3}\left({\bf p}_f\cdot{\bf p}_i\right){\bf S}^{2} 
\label{eq:appb.10} \end{eqnarray}
where the momentum-space tensor-operator $S_{12}$ is defined as
\begin{equation}
    S_{12}(\hat{\bf p})=3 
(\mbox{\boldmath $\sigma$}_1\cdot\hat{\bf p})
(\mbox{\boldmath $\sigma$}_2\cdot\hat{\bf p})
-(\mbox{\boldmath $\sigma$}_1\cdot\mbox{\boldmath $\sigma$}_2)
\label{eq:appb.11} \end{equation}
{}From the formulas given in this appendix the partial wave projections
of the several potential forms, as given in sections \ref{sec:PWA} and
\ref{sec:adPSPS},
can be derived in a straightforward manner. In case of an extra
factor $\left({\bf p}_f\cdot{\bf p}_i\right)$, as occurs for example
in the second line of (\ref{eq:appb.10}), we simply use
the expansion
\begin{equation}
 \left({\bf p}_f\cdot{\bf p}_i\right) V({\bf k}^{2})= p_f p_i
 \sum_{L=0}^{\infty}(2L+1)\tilde{V}_{L}(x) P_{L}(\cos \theta) 
\label{eq:appb.14} \end{equation}
where
\begin{equation}
 \tilde{V}_{L}=\frac{1}{2L+1}\left[(L+1)V_{L+1}+L V_{L-1}\right]\ .
\label{eq:appb.15} \end{equation}
 
\section{ Fourier Transformation Coordinate- to Momentum-space}
\label{app:F}
In this appendix we give an outline of how the potentials in the coordinate
representation can be translated to their momentum space counterparts in a 
direct way. Of course, we utilize the same techniques as described in this 
paper. We treat the more complicated case of the tensor potential. In this 
case the coordinate space potentials are complicated. Nevertheless, we show
explicitly how they are connected with our momentum space representation.

\subsection{ TPS Tensor and Spin-spin Potentials I}        
\label{app:B1}
To appreciate this method in the case of TME-potentials, we consider as
a typical example the potential
\begin{eqnarray}
V(r) &=& \int\frac{d^3k_1}{(2\pi)^3}\ \int\frac{d^3k_2}{(2\pi)^3}\
 e^{i({\bf k_1+k_2})\cdot{\bf r}}\cdot
 \nonumber\\ &\times&
 \left[\mbox{\boldmath $\sigma$}_1\cdot({\bf k}_1\times{\bf k}_2)\right]
 \left[\mbox{\boldmath $\sigma$}_2\cdot({\bf k}_1\times{\bf k}_2)\right]\
 \tilde{F}({\bf k}_1^2) \tilde{G}({\bf k}_2^2)
\label{AppB.1} \end{eqnarray}
This potential, in terms of the Fourier transforms $F(r)$ and $G(r)$ has been
given in \cite{Rij91} and reads
\begin{eqnarray}
V(r) &=& \frac{2}{3}\left[\frac{1}{r^2} F'(r) G'(r)+\frac{1}{r} F'(r) G''(r)
 + \frac{1}{r} F''(r) G'(r)\right]
 \left(\mbox{\boldmath $\sigma$}_1\cdot\mbox{\boldmath $\sigma$}_2\right)\
 \nonumber\\ &&
 +\frac{1}{3}\left[\frac{2}{r^2} F'(r) G'(r)-\frac{1}{r} F'(r) G''(r)
 - \frac{1}{r} F''(r) G'(r)\right]\ S_{12}\ ,
\label{AppB.2} \end{eqnarray}
where $F'(r) \equiv dF(r)/dr$ etc.\\
In seeking the projection on the spinor invariants, we wish to write for the
momentum space counterpart of (\ref{AppB.1}) as
\begin{eqnarray}
\tilde{V}({\bf k}) &=& \int\int\frac{d^3k_1d^3k_2}{(2\pi)^3}\
 \delta({\bf k - k_1 - k_2})\ 
 \tilde{F}({\bf k}_1^2) \tilde{G}({\bf k}_2^2)\
 \left[\mbox{\boldmath $\sigma$}_1\cdot({\bf k}_1\times{\bf k}_2)\right]
 \left[\mbox{\boldmath $\sigma$}_2\cdot({\bf k}_1\times{\bf k}_2)\right]\
 \nonumber\\ & \equiv& \tilde{V}_\sigma({\bf k}) 
 \left(\mbox{\boldmath $\sigma$}_1\cdot\mbox{\boldmath $\sigma$}_2\right)\
 + \tilde{V}_T{\bf k}) \left[
 \left(\mbox{\boldmath $\sigma$}_1\cdot{\bf k}\right)\
 \left(\mbox{\boldmath $\sigma$}_2\cdot{\bf k}\right) - \frac{1}{3}
 \left(\mbox{\boldmath $\sigma$}_1\cdot\mbox{\boldmath $\sigma$}_2\right)\
 {\bf k}^2\right]\ .
\label{AppB.3} \end{eqnarray}
It is now our task to find $\tilde{V}_{\sigma,T}({\bf k})$. To 
proceed, we notice that one can easily see that  
(\ref{AppB.2}) can be written in the form
\begin{eqnarray}
V(r) &\equiv& \frac{2}{3} H_\sigma(r)
 \left(\mbox{\boldmath $\sigma$}_2\cdot\mbox{\boldmath $\sigma$}_2\right)\
 + \frac{1}{3} H_T(r)\ S_{12}\ \ , \ \ {\rm with} \nonumber\\
  H_\sigma(r) &=& \left(\frac{2}{r}+\frac{d}{dr}\right)\ H'(r)\ \ , \ \ {\rm and}\ \
  H_T(r) = \left(\frac{1}{r}-\frac{d}{dr}\right)\ H'(r)\ , 
\label{AppB.4} \end{eqnarray}
where the function $H(r)$ satisfies the equation
\begin{equation}
  \frac{1}{r}\frac{d}{dr} H(r) = \frac{1}{r}\frac{d}{dr} F(r)\cdot
  \frac{1}{r}\frac{d}{dr} G(r)\ .
\label{AppB.5} \end{equation}
To solve the problem posed in this section, it is necessary to find the 
Fourier transform $\tilde{H}({\bf k}^2)$. To solve this problem we exploit the 
following lemma:
\newtheorem{lemma}{Lemma}
\begin{lemma}
If two functions $h(r)$ and $H(r)$ are related by $h(r)=(1/r)dH/dr$, then their fourier
transforms $\tilde{h}({\bf k})$ and $\tilde{H}({\bf k})$ satisfy
\begin{equation}
   h(r) = \frac{1}{r} \frac{d}{dr} H(r) \leftrightarrow 
 \frac{d\tilde{h}({\bf k}^2)}{d{\bf k}^2} = \frac{1}{2} \tilde{H}({\bf k}^2)\ .
\label{AppB.6} \end{equation}
\end{lemma}
Introducing the 'little' functions $h(r),f(r),g(r)$ by
\[
 h(r) \equiv \frac{1}{r}\frac{d}{dr} H(r)\ ,\ f(r) \equiv \frac{1}{r}\frac{d}{dr} F(r)\ ,\
 g(r) \equiv \frac{1}{r}\frac{d}{dr} G(r)\ ,
\] 
and from (\ref{AppB.5}) these satisfy the relation $h(r)= f(r)\cdot g(r)$.

\noindent Next we assume the following generic soft-core forms 
 for $\tilde{F}$ and $\tilde{G}$:
\begin{equation}
 \tilde{F}({\bf k}^2) = \frac{\exp\left[-{\bf k}^2/\Lambda_1^2\right]}{{\bf k}^2+m_1^2}\ ,\
 \tilde{G}({\bf k}^2) = \frac{\exp\left[-{\bf k}^2/\Lambda_2^2\right]}{{\bf k}^2+m_2^2}\ .  
\label{AppB.7} \end{equation}
The solution for $\tilde{f}({\bf k}^2)$ and $\tilde{g}({\bf k}^2)$, using the 
differential equations for the Fourier transfomrs as given in (\ref{AppB.6}), 
is discussed in \cite{RKS91}, Appendix C, and reads
\begin{eqnarray}
 \tilde{f}({\bf k}^2) &=& -\frac{1}{2} \exp\left(m_1^2/\Lambda_1^2\right)\ 
 E_1\left[\left({\bf k}^2+m_1^2\right)/\Lambda_1^2\right]\ , \nonumber\\
 \tilde{g}({\bf k}^2) &=& -\frac{1}{2} \exp\left(m_2^2/\Lambda_2^2\right)\ 
 E_1\left[\left({\bf k}^2+m_2^2\right)/\Lambda_2^2\right]\ ,\
\label{AppB.8} \end{eqnarray}
where $E_1$ is the exponential integral \cite{AS70}.
Then, the Fourier transform of $h(r)=f(r) g(r)$ as follows from (\ref{AppB.5}) is given by 
the convolution
\begin{equation}
 \tilde{h}({\bf k}) = \tilde{f}\star\tilde{g}({\bf k}) = 
 \int \frac{d^3\Delta}{(2\pi)^3}\ \tilde{f}(\mbox{\boldmath $\Delta$}^2)\
 \tilde{g}\left(({\bf k}-\mbox{\boldmath $\Delta$})^2\right)\ .
\label{AppB.9} \end{equation}
With (\ref{AppB.8}) we obtain 
\begin{eqnarray}
 \tilde{h}({\bf k})&=&\frac{1}{4} e^{m_1^2/\Lambda_1^2} e^{m_2^2/\Lambda_2^2} 
 \int \frac{d^3\Delta}{(2\pi)^3}\
 E_1\left[\left(\mbox{\boldmath $\Delta$}^2+m_1^2\right)/\Lambda_1^2\right]\cdot 
 E_1\left[\left(\left({\bf k}-\mbox{\boldmath $\Delta$}\right)^2 
 +m_2^2\right)/\Lambda_2^2\right]\ .           
\label{AppB.10} \end{eqnarray}
Now, the $\Delta$-integral can be performed as follows
\begin{eqnarray}
 && \int \frac{d^3\Delta}{(2\pi)^3}\ E_1\left(\frac{\left(\mbox{\boldmath $\Delta$}^2
 +m_1^2\right)}{\Lambda_1^2}\right)\cdot 
 E_1\left(\frac{\left(\left({\bf k}-\mbox{\boldmath $\Delta$}\right)^2 
 +m_2^2\right)}{\Lambda_2^2}\right) = 
 \int_1^\infty \frac{dt}{t} \int_1^\infty \frac{du}{u}\cdot \nonumber\\
 && \times\int \frac{d^3\Delta}{(2\pi)^3}\ 
 \exp\left[-\frac{\left(\mbox{\boldmath $\Delta$}^2
 +m_1^2\right)}{\Lambda_1^2}t\right]\cdot\exp\left[-
 \frac{\left(\left({\bf k}-\mbox{\boldmath $\Delta$}\right)^2 
 +m_2^2\right)}{\Lambda_2^2} u \right] = \nonumber\\
 && \int_1^\infty \frac{dt}{t} \int_1^\infty \frac{du}{u}
 \exp\left(-\frac{m_1^2}{\Lambda_1^2} t\right)
 \exp\left(-\frac{m_2^2}{\Lambda_2^2} u\right)\cdot \nonumber\\
 && \times \int \frac{d^3\Delta}{(2\pi)^3}\ 
 \exp\left[-\left(\frac{t}{\Lambda_1^2}+\frac{u}{\Lambda_2^2}\right)
 \mbox{\boldmath $\Delta$}^2 + \left(2{\bf k}\cdot\mbox{\boldmath $\Delta$}-{\bf k}^2\right)
 \frac{u}{\Lambda_2^2}\right]\ .            
\label{AppB.11} \end{eqnarray}
The $\Delta$-integral in the last line of (\ref{AppB.11}) has a standard Gaussian form
and is given by
\begin{eqnarray}
 \int \frac{d^3\Delta}{(2\pi)^3}\ \exp\left[ \ldots \vphantom{\frac{A}{A}}\right] &=& 
 \left(4\pi a\right)^{-3/2}\ \exp\left[-\frac{t u/\Lambda_1^2\Lambda_2^2}
 {\left(t/\Lambda_1^2+u/\Lambda_2^2\right)}{\bf k}^2\right]\ 
 ,\ {\rm with}\ a = \frac{t}{\Lambda_1^2} + \frac{u}{\Lambda_2^2}\ .
\label{AppB.12} \end{eqnarray}
In this form, the derivative w.r.t. ${\bf k}^2$ can be taken easily, and we finally obtain 
for $\tilde{H}$ the solution
\begin{eqnarray}
 \widetilde{H}({\bf k}^2) &=& -\frac{1}{2} (4\pi)^{-3/2}
 e^{m_1^2/\Lambda_1^2} e^{m_2^2/\Lambda_2^2} 
 \int_1^\infty \frac{dt}{\Lambda_1^2}\ \int_1^\infty \frac{du}{\Lambda_2^2}\ 
 \left(\frac{t}{\Lambda_1}+\frac{u}{\Lambda_2^2}\right)^{-5/2}
 \cdot \nonumber\\ && \times
 \exp\left(-\frac{m_1^2}{\Lambda_1^2} t\right)
 \exp\left(-\frac{m_2^2}{\Lambda_2^2} u\right)
 \exp\left[-\left(\frac{tu/\Lambda_1^2\Lambda_2^2}{t/\Lambda_1^2+u/\Lambda_2^2}
 \right) {\bf k}^2\right]\ .
\label{AppB.13} \end{eqnarray}
Redefining the variables $t\rightarrow t/\Lambda_1^2$ and 
$u\rightarrow u/\Lambda_2^2$, one can rewrite 
(\ref{AppB.13}) in to the form
\begin{eqnarray}
 \widetilde{H}({\bf k}^2) &=& -\frac{1}{2} (4\pi)^{-3/2}
 e^{m_1^2/\Lambda_1^2} e^{m_2^2/\Lambda_2^2} 
 \int_{t_0}^\infty dt\ \int_{u_0}^\infty du\ \left(t+u \right)^{-5/2} 
 \cdot \nonumber\\ && \times e^{-(m_1^2 t+m_2^2 u)}
 \exp\left[-\left(\frac{tu}{t+u}\right){\bf k}^2\right]\ \  \ (t_0=1/\Lambda_1^2,\
 u_0=1/\Lambda_2^2)\ .
\label{AppB.14} \end{eqnarray}
Notice that this is the same result as obtained in (\ref{AppB.25}), as it should.
 
\end{widetext}

\begin{widetext}
  
\begin{table*}[h]
\caption{Coefficients \protect$\Upsilon^{(0)}_{j,k}$ for the 
         planar (//) adiabatic ps-ps contributions.}                          
\begin{ruledtabular}
\begin{tabular}{c|ccc}  
 &&&\\
    & $\Upsilon_{0}(//)(t,u)$ & $\Upsilon_1(//)(t,u)$ & $\Upsilon_2(//)(t,u)$ \\                       
&&&\\
\hline     
&&&\\
 $\Omega^{(0,//)}_1$ & $\displaystyle{\frac{15}{4}\frac{\sqrt{t+u}}{(t+u)^2}}$ &                    
 $\displaystyle{\frac{1}{2}\left(\frac{t^2-8ut+u^2}{t+u}\right)\frac{\sqrt{t+u}}{(t+u)^2}}$ &    
 $\displaystyle{\frac{t^2u^2}{(t+u)^2}\frac{\sqrt{t+u}}{(t+u)^2}}$  \\[3mm]      
 $\Omega^{(0,//)}_2$ & ---  &$\displaystyle{-\frac{1}{3}\frac{\sqrt{t+u}}{(t+u)}}$ &--- \\[3mm] 
 $\Omega^{(0,//)}_3$ & $+\displaystyle{\frac{1}{2}\frac{\sqrt{t+u}}{(t+u)}}$ &--- &--- \\[3mm]    
\end{tabular}
\end{ruledtabular}
\label{table1}
\end{table*}
 
\begin{table*}[h]
\caption{Coefficients \protect$\Upsilon^{(0)}_{j,k}$ for the 
         crossed (X) adiabatic ps-ps contributions.}      
\begin{ruledtabular}                    
\begin{tabular}{c|ccc}  
 &&&\\
    & $\Upsilon_{0}(X)(t,u)$ & $\Upsilon_1(X)(t,u)$ & $\Upsilon_2(X)(t,u)$ \\                       
 &&&\\
\hline     
 &&&\\
 $\Omega^{(0,X)}_1$ & $-\displaystyle{\frac{15}{4}\frac{\sqrt{t+u}}{(t+u)^2}}$ &                    
 $\displaystyle{-\frac{1}{2}\left(\frac{t^2-8ut+u^2}{t+u}\right)\frac{\sqrt{t+u}}{(t+u)^2}}$ &      
 $\displaystyle{-\frac{t^2u^2}{(t+u)^2}\frac{\sqrt{t+u}}{(t+u)^2}}$  \\[3mm]        
 $\Omega^{(0,X)}_2$ & ---  &$\displaystyle{-\frac{1}{3}\frac{\sqrt{t+u}}{(t+u)}}$ &--- \\[3mm]       
 $\Omega^{(0,X)}_3$ & $+\displaystyle{\frac{1}{2}\frac{\sqrt{t+u}}{(t+u)}}$ &--- &--- \\[3mm]       
\end{tabular}
\end{ruledtabular}
\label{table2}
\end{table*}
 
\begin{table*}[h]
\caption{Coefficients \protect$\Upsilon^{(1a)}_{j,k}$ for the 
         planar (//) non-adiabatic ps-ps contributions.}                       \begin{ruledtabular}   
\begin{tabular}{c|cccc}  
 &&&&\\
    & $\Upsilon_0^{(1a)}(//)(t,u)$ & $\Upsilon_1^{(1a)}(//)(t,u)$ & 
      $\Upsilon_2^{(1a)}(//)(t,u)$ & $\Upsilon_3^{(1a)}(//)(t,u)$ \\                       
 &&&&\\
\hline     
 &&&&\\
 $\Omega^{(1a,//)}_1$ & $\displaystyle{-\frac{105}{8}\frac{1}{(t+u)^3}}$ &                    
 $\displaystyle{-\frac{15}{4}\left(\frac{t^2-5ut+u^2}{(t+u)^4}\right)}$ &
 $\displaystyle{\frac{3}{2}\left(\frac{t^2-5ut+u^2}{(t+u)^5}\right)}$ &              
 $\displaystyle{\frac{t^3u^3}{(t+u)^6}}$  \\[3mm]        
 $\Omega^{(1a,//}_2$ & ---  &$\displaystyle{\frac{5}{6}\frac{1}{(t+u)^2}}$ &      
 $\displaystyle{-\frac{1}{3}\frac{tu}{(t+u)^3}}$ &--- \\[3mm]       
 $\Omega^{(1a,//)}_3$ & $-\displaystyle{\frac{5}{4}\frac{1}{(t+u)^2}}$ &       
 $\displaystyle{\frac{1}{2}\frac{tu}{(t+u)^3}}$ &---&---\\[3mm]       
 $\Omega^{(1a,//)}_4$ & $-\displaystyle{\frac{1}{(t+u)^2}}$ &       
 $\displaystyle{\frac{2tu}{(t+u)^3}}$ &--- &---\\[3mm]       
\end{tabular}
\end{ruledtabular}
\label{table3}
\end{table*}
 
\begin{table*}
\caption{Coefficients \protect$\Upsilon^{(1a)}_{j,k}$ for the 
         crossed (X) non-adiabatic ps-ps contributions.}                       \begin{ruledtabular}   
\begin{tabular}{c|cccc}  
 &&&\\
    & $\Upsilon_0^{(1a)}(X)(t,u)$ & $\Upsilon_1^{(1a)}(X)(t,u)$ & 
      $\Upsilon_2^{(1a)}(X)(t,u)$ & $\Upsilon_3^{(1a)}(X)(t,u)$ \\                       
 &&&&\\
\hline
 &&&&\\
 $\Omega^{(1a,X)}_1$ & $\displaystyle{+\frac{105}{4}\frac{1}{(t+u)^3}}$ &                    
 $\displaystyle{+\frac{15}{2}\left(\frac{t^2-5ut+u^2}{(t+u)^4}\right)}$ &
 $\displaystyle{-3\left(\frac{t^2-5ut+u^2}{(t+u)^5}\right)}$ &              
 $\displaystyle{-2\frac{t^3u^3}{(t+u)^6}}$  \\[3mm]        
 $\Omega^{(1a,X}_2$ & ---  &$\displaystyle{\frac{5}{3}\frac{1}{(t+u)^2}}$ &       
 $\displaystyle{-\frac{2}{3}\frac{tu}{(t+u)^3}}$ &--- \\[3mm]       
 $\Omega^{(1a,//)}_3$ & $-\displaystyle{\frac{5}{2}\frac{1}{(t+u)^2}}$ &       
 $\displaystyle{\frac{tu}{(t+u)^3}}$ &---&--- \\[3mm]       
\end{tabular}
\end{ruledtabular}
\label{table4}
\end{table*}
 
\begin{table*}
\caption{Coefficients \protect$\Upsilon^{(1b)}_{j,k}$ for the 
         pseudovector-vertex correction ps-ps contributions.}                  \begin{ruledtabular}        
\begin{tabular}{c|ccc}  
 &&&\\
    & $\Upsilon_{0}^{(1b)}(t,u)$ & $\Upsilon_1^{(1b)}(t,u)$ & 
      $\Upsilon_2^{(1b)}(t,u)$ \\                       
 &&&\\
\hline
 &&&\\
 $\Omega^{(1b,X)}_1$ & $-\displaystyle{\frac{15}{2}\frac{1}{(t+u)^2}}$ &                    
 $-\displaystyle{\frac{5}{2}\left(\frac{t^2-2ut+u^2}{t+u}\right)\frac{1}{(t+u)^2}}$ &             
 $ \displaystyle{\frac{t^2+u^2}{(t+u)^2}\frac{tu}{(t+u)^2}}$  \\[3mm]        
 $\Omega^{(1b,X)}_4$ & $\displaystyle{-\frac{1}{t+u}}$ & 
 $\displaystyle{\frac{2tu}{(t+u)}}$ &--- \\[3mm]       
\end{tabular}
\end{ruledtabular}
\label{table5}
\end{table*}
 
\begin{table*}
\caption{Coefficients \protect$\Upsilon^{(1c)}_{j,k}$ for the 
         off-shell corrections TMO-graphs ps-ps contributions.}                \begin{ruledtabular}          
\begin{tabular}{c|ccc}  
 &&&\\
    & $\Upsilon_{0}^{(1c)}(t,u)$ & $\Upsilon_1^{(1c)}(t,u)$ & 
      $\Upsilon_2^{(1c)}(t,u)$ \\                       
 &&&\\
\hline
 &&&\\
 $\Omega^{(1c,TMO)}_1$ & $\displaystyle{-\frac{15}{4}\frac{1}{(t+u)^2}}$ &                    
 $\displaystyle{-\frac{5}{4}\left(\frac{t^2-2ut+u^2}{t+u}\right)\frac{1}{(t+u)^2}}$ &           
 $\displaystyle{\frac{1}{2}\frac{t^2+u^2}{(t+u)^2}\frac{tu}{(t+u)^2}}$  \\[3mm]        
 $\Omega^{(1c,TMO)}_4$ & $\displaystyle{-\frac{5}{2}\frac{1}{t+u}}$ & 
 $\displaystyle{\frac{tu}{(t+u)}}$ &--- \\[3mm]       
\end{tabular}
\end{ruledtabular}
\label{table6}
\end{table*}

\twocolumngrid
\end{widetext}


\begin{thebibliography}{99}
\bibitem{footnote}     
 As pointed out in \cite{RS96a}, the operators $\widetilde{O}^{(//,X)}_{\alpha\beta}$
 contain the contributions from all graphs, including the graphs with 
 meson $1$ and meson $2$ are interchanged. To avoid overcounting, in case 
 of identical mesons, like $\pi\pi$ or $\eta\eta$ these operators must be
 divided by 2. The results in the Tables of this paper are given for non-identical mesons.
\bibitem{Somers}
 The explicit relation between our $9j$-symbols and those of
 \cite{Edmonds57} eq.~(6.4.4) is
 \begin{eqnarray} \left[ \begin{array}{ccc}
        j_{11} & j_{12} & j_{13} \\[0.2cm]
        j_{21} & j_{22} & j_{23} \\[0.2cm]
        j_{31} & j_{32} & j_{33} \\
 \end{array} \right]
 & = &
 \left[(2j_{13}+1)(2j_{31}+1)(2j_{23}+1)(2j_{32}+1)\right]^{1/2}\
 \nonumber \\[0.2cm]  && \times (-)^{j_{31}+j_{13}-j_{32}-j_{23}}\cdot
      \left\{ \begin{array}{ccc}
        j_{11} & j_{12} & j_{13} \\[0.2cm]
        j_{21} & j_{22} & j_{23} \\[0.2cm]
        j_{31} & j_{32} & j_{33} \\
 \end{array} \right\}  \nonumber
 \end{eqnarray}
\end{thebibliography}
\end{document}